\documentclass[%
 reprint,
 amsmath,amssymb,
 aps,prl,superscriptaddress
]{revtex4-2}

\usepackage{graphicx}
\usepackage{dcolumn}
\usepackage{bm}
\usepackage{xcolor}

\begin{document}


\title{Discovery of an electronic crystal in a cuprate Mott insulator}

\author{Mingu Kang}
\thanks{These authors contributed equally}
\affiliation{Department of Physics, Massachusetts Institute of Technology; Cambridge, MA 02139, USA}
\affiliation{Center for Complex Phase of Materials, Max Planck POSTECH Korea research Initiative; Pohang 790-884, Republic of Korea}
\author{Charles Zhang}
\thanks{These authors contributed equally}
\affiliation{Department of Physics, University of Toronto, Toronto, Ontario M5S1A7, Canada}
\author{Enrico Schierle}
\affiliation{Helmholtz-Zentrum Berlin für Materialien und Energie; Albert-Einstein Strabe 15, D-12489 Berlin, Germany}
\author{Stephen McCoy}
\affiliation{Material Science and Engineering, University of California, Riverside; Riverside, California 92521, USA}
\affiliation{Department of Electrical Engineering, University of California, Riverside, Riverside; California 92521, USA}
\author{Jiarui Li} 
\affiliation{Department of Physics, Massachusetts Institute of Technology; Cambridge, MA 02139, USA}
\author{Ronny Sutarto} 
\affiliation{Canadian Light Source; Saskatoon, Saskatchewan S7N 2V3, Canada}
\author{Feizhou He}
\affiliation{Canadian Light Source; Saskatoon, Saskatchewan S7N 2V3, Canada}
\author{Andreas Suter}
\affiliation{Laboratory for Muon-Spin Spectroscopy, Paul Scherrer Institute, CH-5232 Villigen, Switzerland}
\author{Thomas Prokscha}
\affiliation{Laboratory for Muon-Spin Spectroscopy, Paul Scherrer Institute, CH-5232 Villigen, Switzerland}
\author{Zaher Salman}
\affiliation{Laboratory for Muon-Spin Spectroscopy, Paul Scherrer Institute, CH-5232 Villigen, Switzerland}
\author{Eugen Weschke}
\affiliation{Helmholtz-Zentrum Berlin für Materialien und Energie; Albert-Einstein Strabe 15, D-12489 Berlin, Germany}
\author{Shane Cybart}
\affiliation{Material Science and Engineering, University of California, Riverside; Riverside, California 92521, USA}
\affiliation{Department of Electrical Engineering, University of California, Riverside, Riverside; California 92521, USA}
\author{John Y. T. Wei}
\affiliation{Department of Physics, University of Toronto, Toronto, Ontario M5S1A7, Canada}
\author{Riccardo Comin}
\email{rcomin@mit.edu}
\affiliation{Department of Physics, Massachusetts Institute of Technology; Cambridge, MA 02139, USA}
\date{\today}

\begin{abstract}
Copper oxide high temperature superconductors universally exhibit multiple forms of electronically ordered phases that break the native translational symmetry of the CuO$_2$ planes. The interplay between these orders and the superconducting ground state, as well as how they arise through doping a Mott insulator, is essential to decode the mechanisms of high-temperature superconductivity. Over the years, various forms of electronic liquid crystal phases -- including charge/spin stripes and incommensurate charge-density-waves (CDWs) -- were found to emerge out of a correlated metallic ground state in underdoped cuprates. Early theoretical studies also predicted the emergence of a Coulomb-frustrated 'charge crystal' phase in the very lightly-doped, insulating limit of the CuO$_{2}$ planes. This charge crystal phase was observed only recently at the surface of Bi$_2$Sr$_2$CaCu$_{2}$O$_{8+y}$, raising new questions on whether such a state may also exist in the bulk and thus be a common instability of cuprates. Here, we use resonant X-ray scattering, electron transport, and muon spin rotation measurements to fully resolve the electronic and magnetic ground state and search for signatures of charge order in very lightly hole-doped cuprates from the 123 family \textit{R}Ba$_2$Cu$_3$O${}_{7 - \delta}$ (\textit{R}BCO; \textit{R}: Y or rare earth). X-ray scattering data from \textit{R}BCO films reveal a breaking of translational symmetry more pervasive than was previously known, extending down to the insulating and magnetically ordered Mott limit. The ordering vector of this charge crystal state is linearly connected to the charge-density-waves of underdoped \textit{R}BCO, suggesting that the former phase is a precursor to the latter as hole doping is increased. Most importantly, the coexistence of charge and spin order in \textit{R}BCO suggests that this electronic symmetry-breaking state is common to the CuO$_2$ planes in the very lightly-doped regime. These findings bridge the gap between the Mott insulating state and the underdoped metallic state, and underscore the prominent role of Coulomb-frustrated electronic phase separation among all cuprates.
\end{abstract}

\maketitle


\section{\label{sec:level1} I. Introduction}


Following the discovery of high-temperature superconductivity \cite{bednorz_possible_1986}, early theoretical studies proposed that multiple forms of electronic symmetry-breaking exist in lightly-doped Mott insulators \cite{zaanen_charged_1989, machida_magnetism_1989, poilblanc_charged_1989, emery_phase_1990}. These theoretical predictions were confirmed by the discovery of charge and spin stripes in La$_{1.48}$Nd$_{0.4}$Sr$_{0.12}$CuO$_4$ \cite{tranquada_stripes_1995}. During the last decade,multiple forms of these symmetry-breaking states have been discovered in all cuprate high-temperature superconductors \cite{Ghiringhelli2012,Chang2012,Blanco-Canosa2014, Comin2014,DaSilvaNeto2014,Tabis2014,DasilvaNeto2015,DaSilvaNeto2016,Kang2019,Comin2016}.
In many cases, these emergent phases are intertwined with superconductivity \cite{kivelson_how_2003, tranquada_exploring_2015, fradkin_colloquium_2015} in ways that have been investigated from multiple perspectives \cite{Kivelson1998,Vojta1999,Sachdev2013,Davis2013,Lee2014,Wang2014,DallaTorre2015,Duong2017}.

The driving mechanisms behind electronic order in the CuO$_2$ planes have been the subject of longstanding questions that are essential for understanding of quantum matter and high-temperature superconductivity in cuprates \cite{keimer_quantum_2015}. However, a consensus on the origin of electronic order has not formed fully, partly because of key phenomenological differences between the cuprate families of YBa$_{2}$Cu$_{3}$O$_{y}$ (123) Bi$_2$Sr$_2$CaCu$_{2}$O$_{8+y}$ (2212), and La$_{2}$CuO$_{4}$ (214). In the latter, charge order cohabits with static spin order to form electronic stripes characterized by Coulomb-frustrated real-space phase separation and a spatial periodicity directly correlated with the hole density. However, this 'real-space Coulomb frustration' scenario has not found wide applicability in other cuprate families, because of the competing hypothesis of a 'momentum-space instability' involving scattering between itinerant states mediated by strong antiferromagnetic fluctuations.

\begin{figure*}
\includegraphics[width=2.0 \columnwidth]{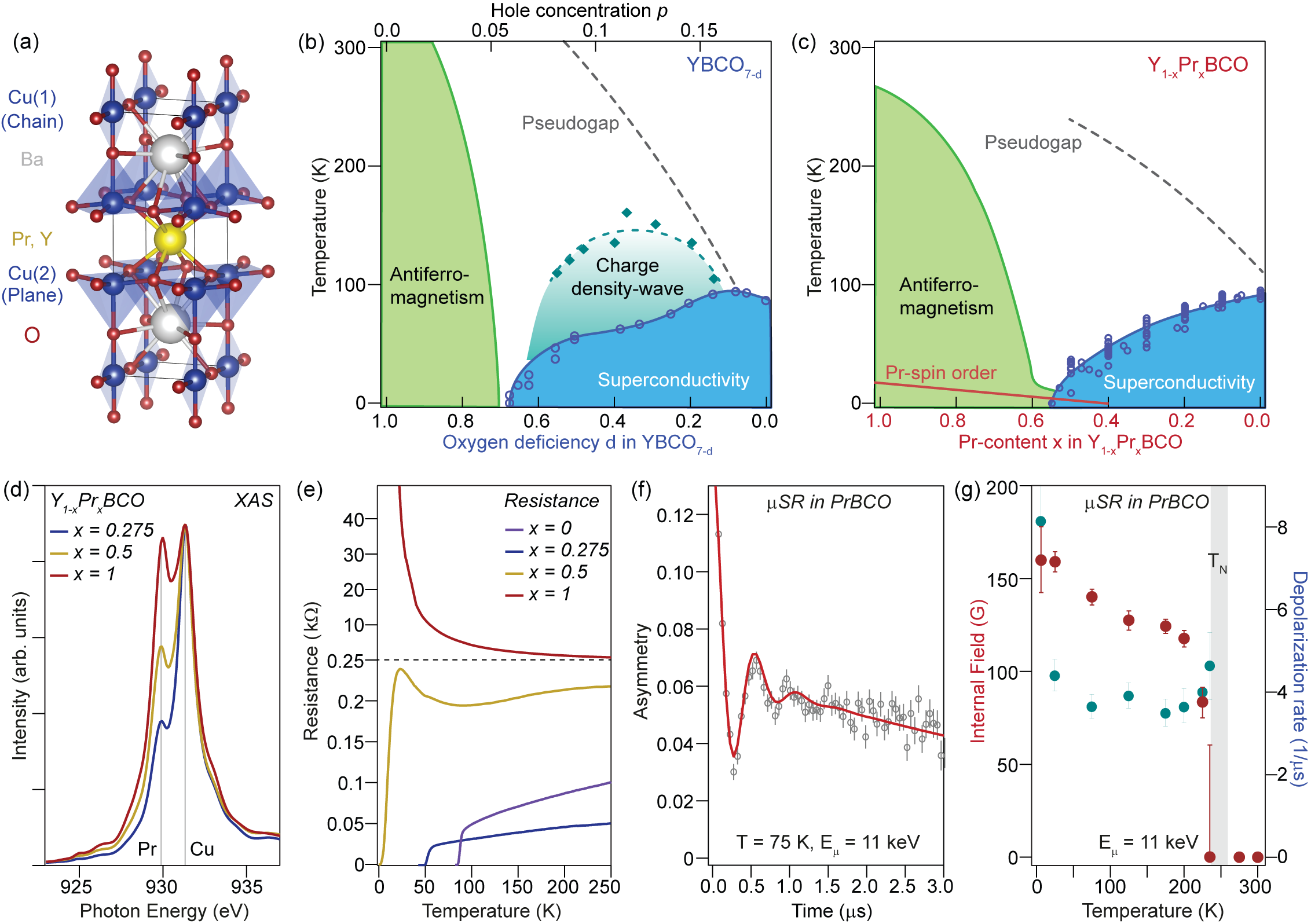}
\caption{\label{fig:1} Characterization of Pr-substituted YBCO and its phase diagram. (a) Crystal structure of Y$_{1-x}$Pr$_{x}$BCO containing a CuO$_{2}$-Y,Pr-CuO$_{2}$ bilayer. (b,c) Phase diagram of YBCO$_{7-\delta}$ (b) and Y$_{1-x}$Pr$_{x}$BCO (c). Doping dependent data points for superconductivity in YBCO$_{7-\delta}$, superconductivity in Y$_{1-x}$Pr$_{x}$BCO, and charge order in YBCO$_{7-\delta}$ are adapted from Refs. \cite{Blanco-Canosa2014,Radousky1992,Liang2006}. (d) X-ray absorption spectra of Y$_{1-x}$Pr$_{x}$BCO films. Two main peaks located at 931.3 eV and 930 eV corresponds to the Cu-$L_{3}$ and Pr-$M_{5}$ absorption edges respectively. (e) Resistance of Y$_{1-x}$Pr$_{x}$BCO films as a function of temperature (note the use of two different linear scales on the $y$-axis above and below 250 $\Omega$). (f) Asymmetry spectrum of muon spin rotation data from PrBCO films at 75 K. (g) Temperature-dependent internal field and depolarization rate. Each data point is obtained by fitting the asymmetry spectra taken at different temperatures. Grey-shaded bars indicate the onset of magnetic order at $T_{N} \approx$ 250 K. A magnetic volume fraction close to 100\% has been deduced from weak transverse field experiments (${B}_{ext}=75 G$) measurements (not shown), indicating that the film is fully magnetic.}
\end{figure*}

A clear discriminant between these two frameworks is the existence of an electronic crystal phase in the very lightly hole-doped and insulating CuO$_2$, before the onset of superconductivity. This state has been predicted under the real-space scenario \cite{kivelson_electronic_1998} but it is fundamentally incompatible with the momentum-space scenario which requires a correlated metallic state with low-energy quasiparticles and a Fermi surface. A recent microscopy study of Bi-based cuprates has found periodic spatial modulations in the tunneling conductance of an insulating Bi-2212 surface, providing fresh evidence for an electronic crystal state \cite{zhao_charge-stripe_2019}. This study demonstrates that the phenomenology of electronic orders in underdoped Bi-based cuprates needs to be understood on the basis of the real-space Coulomb frustration picture. If confirmed in the family of Y-based cuprates, these findings would have a significant impact on our description of electronic orders in the CuO$_2$ planes across all cuprates.

To address this question, we turn our attention to the 123 family of $R$Ba$_{2}$Cu$_{3}$O$_{7-\delta}$ compounds (\textit{R}BCO; \textit{R}: Y or rare earth). Among cuprate high-temperature superconductors, the \textit{R}BCO family is one of the most widely studied, owing to its low degree of electronic disorder that is enabled by sourcing the carrier doping far away from the CuO$_{2}$ planes. This possibility is uniquely enabled by the crystal structure of \textit{R}BCO, with the CuO chain layer providing a charge reservoir for the CuO$_{2}$ planes, which form a bilayer around the spacer rare earth layer [Fig. 1(a)]. To sidestep challenges in the control of oxygen stoichiometry in the chain layer at very low hole doping levels, we have taken an alternative route of using Pr-substitution at the Y site in nearly fully-oxygenated ($\delta \ll 1$) samples of (Y,Pr)Ba$_{2}$Cu$_{3}$O$_{7}$ (Y$_{1-x}$Pr$_{x}$BCO${}_{7 - \delta}$), taking cues from previous studies of doping control and suppression of superconductivity in these systems [see Fig. 1(c) for the phase diagram of Y$_{1-x}$Pr$_{x}$BCO] \cite{Radousky1992,Akhavan2002,Soderholm1987}. In Y$_{1-x}$Pr$_{x}$BCO, the concentration of mobile carriers in the CuO$_2$ planes is controlled via rare earth substitution (Pr-for-Y) using the known ability of low-energy Pr $4f$ orbitals to effectively sequester holes from Cu-${3d}_{x^2 - y^2}$-O$_{{2p}_{\sigma}}$ states \cite{neumeier_hole_1989,sun_electron_1994}. The fully-substituted, end compound PrBCO is known to be insulating and magnetic [Fig. 1(c)], therefore providing an ideal platform to study electronic orders down to the Mott limit.



Using resonant soft X-ray scattering (RXS) at Cu-$L_{3}$ (2$p \rightarrow$ 3$d$) and Pr-$M_{5}$ (3$d \rightarrow$ 4$f$) edges, we reveal an electronically modulated state across the extended phase diagram of Y$_{1-x}$Pr$_{x}$BCO, from the underdoped superconducting regime ($x$ = 0.275 and 0.5) to the Mott limit ($x$ = 1). In PrBCO, the observation of a Cu-site density wave in the absence of a Fermi surface and its coexistence with static Neel order crucially underscores the need for Coulomb frustration to support charge order in \textit{R}BCO.

\section{\label{sec:level1}{II. Charge order \\in magnetic insulating P\lowercase{r}B\lowercase{a}$_{2}$C\lowercase{u}$_{3}$O$_{7-\delta}$}}

We first present the basic characterization of Y$_{1-x}$Pr$_{x}$BCO films with Pr content $x$ = 0.275, 0.5, and 1. X-ray absorption spectra are shown in Figure 1(d). Two prominent absorption peaks are observed at 931.3 eV and 930.0 eV, corresponding to the Cu-$L_{3}$ and Pr-$M_{5}$ resonances, respectively. The intensity of the Pr-$M_{5}$ peak grows in accordance with higher Pr content $x$. Transport properties of the Y$_{1-x}$Pr$_{x}$BCO films are also displayed in Fig. 1(e). The $T_{c}$ of pristine YBCO at this oxygen content is $\approx$ 90 K in accordance with prior reports \cite{Blanco-Canosa2014,Frano2016,Liang2006}. With Pr substitution, $T_{c}$ is reduced to 55 K at $x$ = 0.275 [Fig. 1(e), $T_{c}$ is estimated from the midpoint of the transition], and is further suppressed to 12 K for $x$ = 0.5. In the latter sample, an upturn in resistivity is manifested at low temperatures, signaling the proximity to a metal-insulator transition. Finally, $x$ = 1 PrBCO behaves as a strong insulator with its resistivity diverging at low temperature. We also performed zero field muon spin rotation (ZF-$\mu$SR) measurements to characterize the magnetic properties of the end compound PrBCO. As shown in Fig. 1(f), the ZF-$\mu$SR time spectrum at 75 K exhibits time oscillations that are indicative of the presence of static magnetic order in the PrBCO film. By fitting the ZF-$\mu$SR spectra using a standard function (see Appendix for detail), we extract the internal field and depolarization rate as a function of temperature in Fig. 1(g). The internal field shows a sharp onset of static spin order at $T_{N} \approx$ 250 K, where the decay rate also peaks. The latter shows a second rise below 25 K due to the additional ordering of Pr magnetic moments. Overall, the electronic and magnetic properties of the Y$_{1-x}$Pr$_{x}$BCO thin films studied here are quantitatively consistent with previous reports \cite{Soderholm1987,Radousky1992,sun_electron_1994,wojek_2012}, confirming that Pr-substitution effectively underdopes the CuO$_2$ planes all the way to the magnetically ordered, insulating limit [Fig. 1(c)].

\begin{figure}
\includegraphics[width=1.0 \columnwidth]{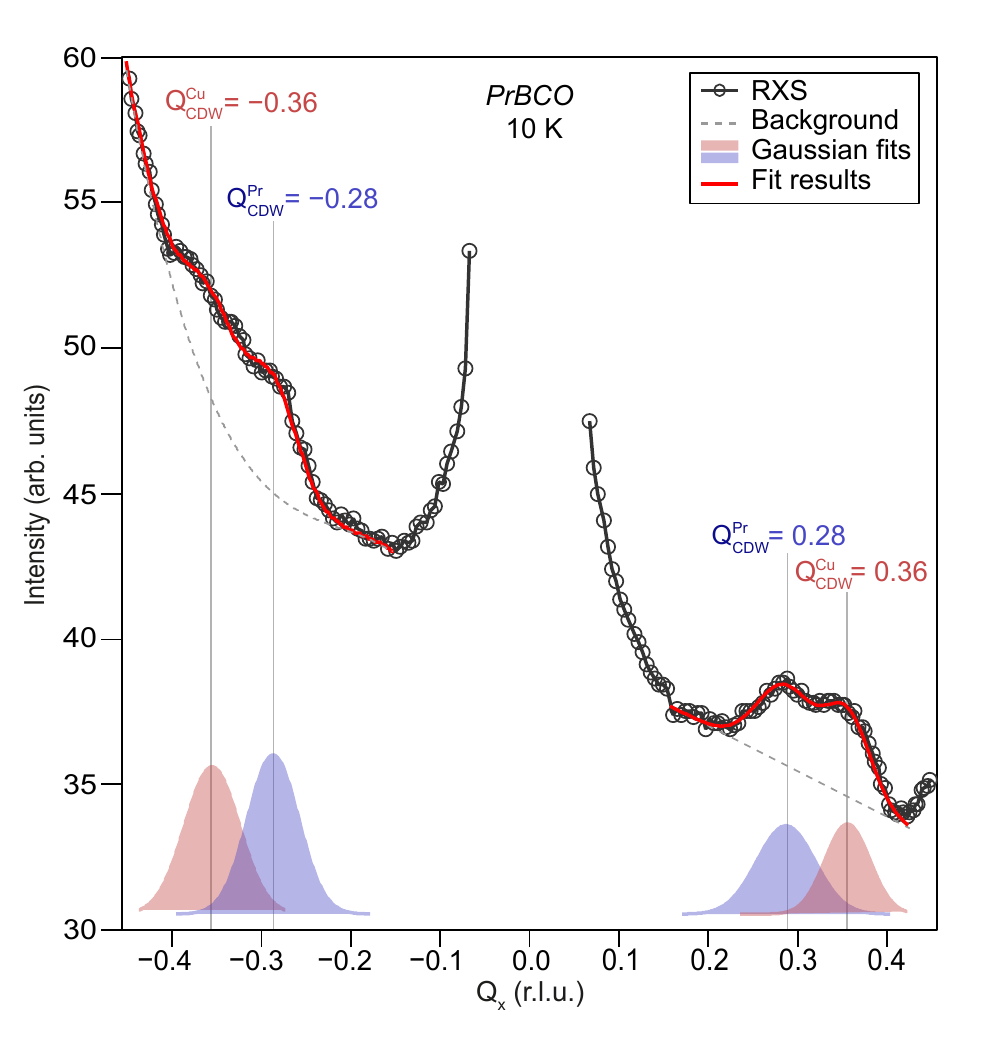}\
\caption{\label{fig:epsart} Charge order in PrBCO. Wide momentum-range RXS scans on PrBCO along the Cu-O bond direction ($Q_{x}$) obtained at the Cu-$L_{3}$ edge. The double peak structure is marked by red solid lines. The grey-dashed lines are used to model the fluorescence background, while the red- and blue-shaded areas are fitted Gaussian peaks.}
\end{figure}

\begin{figure*}
\includegraphics[width=1.5 \columnwidth]{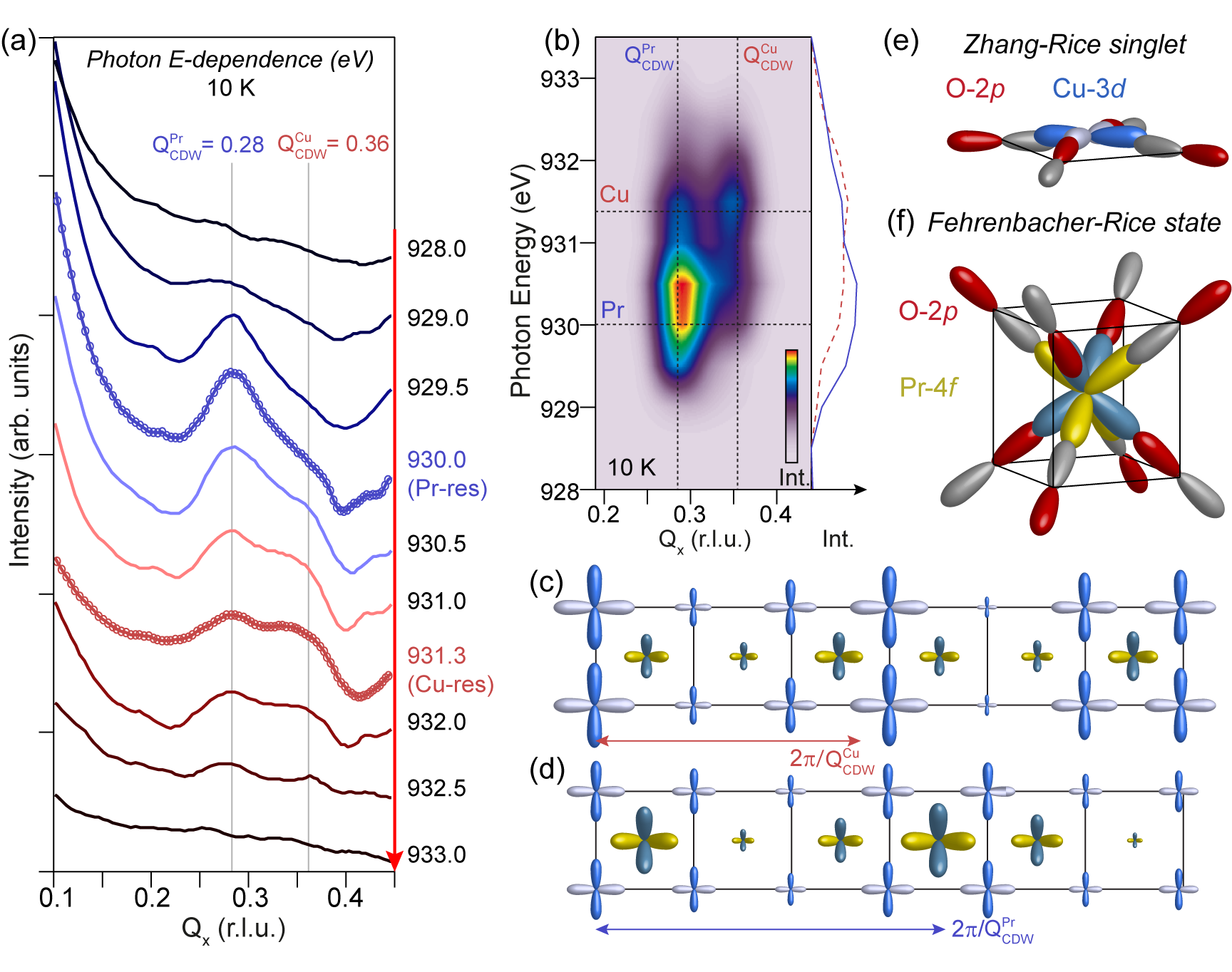}
\caption{\label{fig:2} Resonant profile of charge order and hybridization between Pr-4$f$ and CuO$_{2}$ orbitals. (a) Photon energy-dependence of the X-ray scattering data (momentum scans) for positive $Q_{x}$. The curves are vertically shifted for clarity. Two curves with data points marked with open circles highlight the RXS scans at the peak of the Pr- and Cu-resonances. (b) Intensity map representing the resonant behavior of charge order after subtraction of the fluorescence background. The red-dashed and blue-solid curves on the right show the resonant profile at $Q_{\mathrm{CDW}}^{\mathrm{Cu}}$ and $Q_{\mathrm{CDW}}^{\mathrm{Pr}}$ respectively. (c,d) Schematics of the two density-waves with different Cu-3$d$ and Pr-4$f$ orbital contents. (e) Schematic of the Zhang-Rice singlet state formed upon hybridization of planar Cu-3$d_{x2-y2}$ orbital and O-2$p$ hole states. (f) Schematic of the Fehrenbacher-Rice state formed upon hybridization of Pr-4$f_{z(x2-y2)}$ orbital and surrounding O-2$p$ hole states.}
\end{figure*}

Figure 2 shows a RXS scan taken in PrBCO over a broad range of in-plane momentum along the Cu-O bond direction ($Q_{x}$). On top of a smooth background due to fluorescence and the tail of a specular reflection near $Q_{x}$ = 0, a two-peak structure appears symmetrically at positive and negative momentum transfers between $\vert Q_{x}\vert \approx$ 0.2 and 0.4 r.l.u.. Gaussian fits of the RXS profile reveal that the diffraction signatures are composed of two distinct peaks centered at 0.36 and 0.28 r.l.u.. The full-width-at-half-maximum in both cases is estimated to be $\approx$ 0.06-0.08 r.l.u., reflecting the presence of short-range charge correlations with a spatial coherence length $\xi \approx$ 4-5 unit cells or $\approx$ 15-20 \AA. Both diffraction peaks originate from ordering of the charge degrees of freedom as confirmed by the photon polarization analysis (see Fig. S1 in Supplemental Material) \cite{Ghiringhelli2012, SI}. Additional azimuthal angle-dependent experiments indicate that the double-peak structure appears only along the Cu-O bond direction (see Fig. S2 in Supplemental Material) \cite{SI}. The correlation length, polarization dependence, and ordering axis of these diffraction signatures are all consistent with the characteristics of charge order in other cuprates. However, the coexistence of Cu-$3d$ charge order with static Cu-$3d$ spin order departs from all previous studies, signaling the emergence of a new state.

To understand these initial observations, we performed RXS experiments as a function of photon energy across the Cu-$L_{3}$ (931.3 eV) and Pr-$M_{5}$ (930.0 eV) edges. As shown in Fig. 3(a), the two main diffraction peaks display a surprising evolution with photon energy. When X-rays are tuned more than 2 eV away from the Pr and Cu resonances, the peaks vanish (see the profiles at 928.0 and 933.0 eV for examples), indicating that the diffraction signatures arise from charge modulations within the CuO$_{2}$-Pr-CuO$_{2}$ bilayer. Most importantly, and as can be seen in the intensity color map of Fig. 3(b), the first diffraction peak, hereafter termed $Q_{\mathrm{CDW}}^{\mathrm{Pr}} \approx $\,0.28 r.l.u., resonates at the Pr-edge, while the second peak $Q_{\mathrm{CDW}}^{\mathrm{Cu}} \approx $\,0.36 r.l.u. is most prominent at the Cu-edge (931.3 eV).

\begin{figure*}
\includegraphics[width=1.3 \columnwidth]{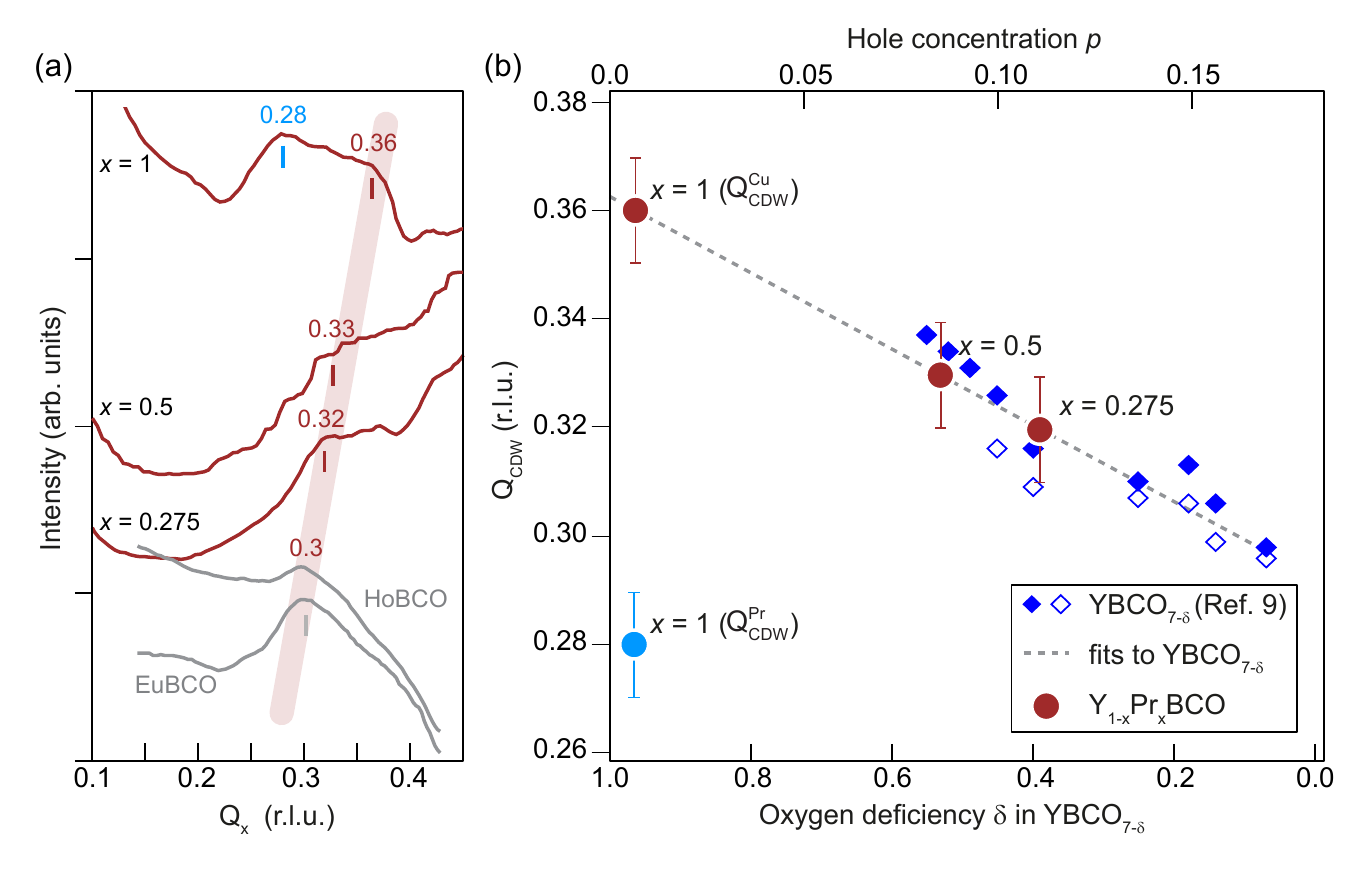}
\caption{\label{fig:2} Charge order as a function of doping. (a) Doping-dependent RXS spectra from Y$_{1-x}$Pr$_{x}$BCO films ($x$ = 0.275, 0.5, 1). RXS spectra from nearly optimally doped films of HoBCO and EuBCO (our 'control' samples) are also displayed. (b) Evolution of the charge order wavevectors in the \textit{R}BCO family. Open and filled blue diamonds represent $Q_{CDW}$ of YBCO$_{7-\delta}$ along the $a$- and $b$-crystallographic axes, respectively (adopted from Ref. \cite{Blanco-Canosa2014}). The relationship between $Q_{CDW}$ and $p$ is obtained from the fit to the YBCO$_{7-\delta}$ series (grey-dashed line), from which we evaluate the hole-doping levels of Y$_{1-x}$Pr$_{x}$BCO.}
\end{figure*}

\section{\label{sec:level1} III. Doping dependence of the diffraction data}

X-ray scattering data indicate the presence of two electronic modulations. The $Q_{\mathrm{CDW}}^{\mathrm{Pr}}$ peak can be ascribed to a spatial modulation of the Pr-4$f$ orbital occupation [see schematics in Fig. 3(d)]. We interpret this signal as arising from Pr-4$f$-O$_{{2p}_{\pi}}$ hybridized valence states [Fig. 3(f)] which have been proposed within the Fehrenbacher-Rice model \cite{Fehrenbacher1993}.

On the other hand, the $Q_{\mathrm{CDW}}^{\mathrm{Cu}}$ peak corresponds to a periodic modulation of the occupation of Cu-${3d}_{x^2 - y^2}$ orbitals [Fig. 3(c,e)] and thus is closely related to the CDW order observed in YBCO. This connection is validated by comparing the CDW signatures of PrBCO with those we measured in 'control' samples from the \textit{R}BCO family. Namely, in superconducting EuBCO and HoBCO films with near-optimal oxygen content, the RXS profiles (See Supplemental Materials Fig. S4) resonate at the same energy but have a very different ordering vector $Q_{\mathrm{CDW}}^{\mathrm{Cu}} \approx$ 0.3. For samples at intermediate doping within the (Y,Pr)BCO series, we find $Q_{\mathrm{CDW}}^{\mathrm{Cu}} \approx$ 0.32 r.l.u. and 0.33 r.l.u. for $x$ = 0.275 and $x$ = 0.5, respectively. This progressive increase of the ordering vector as doping is reduced can be understood on the basis of the well-established dependence of the ordering vector on hole doping, which has been previously used to 'gauge' the doping level in oxide heterostructures \cite{Frano2016}. Here, we use this approach to infer the carrier filling of the CuO$_2$ planes of PrBCO from the relationship between $Q_{\mathrm{CDW}}^{\mathrm{Cu}}$ and hole doping ($p$) in the YBCO$_{7-\delta}$ series [Fig. 4(b); data from Ref.\,\cite{Blanco-Canosa2014}]. Using $Q_{\mathrm{CDW}}^{\mathrm{Cu}}$ = 0.32 and 0.33 r.l.u. for $x$ = 0.275 and 0.5 and $Q_{\mathrm{CDW}}^{\mathrm{Cu}}$ = 0.36 r.l.u. for $x$ = 1 yields hole concentrations $p$ = 0.11, 0.085, and 0.01, respectively. This latter estimate indicates that the end compound PrBCO is in the very lightly hole-doped Mott limit, in excellent accordance with the observed insulating and magnetic ground state (with high Neel temperature $T_{N}$).


\begin{figure*}
\includegraphics[width=2.0 \columnwidth]{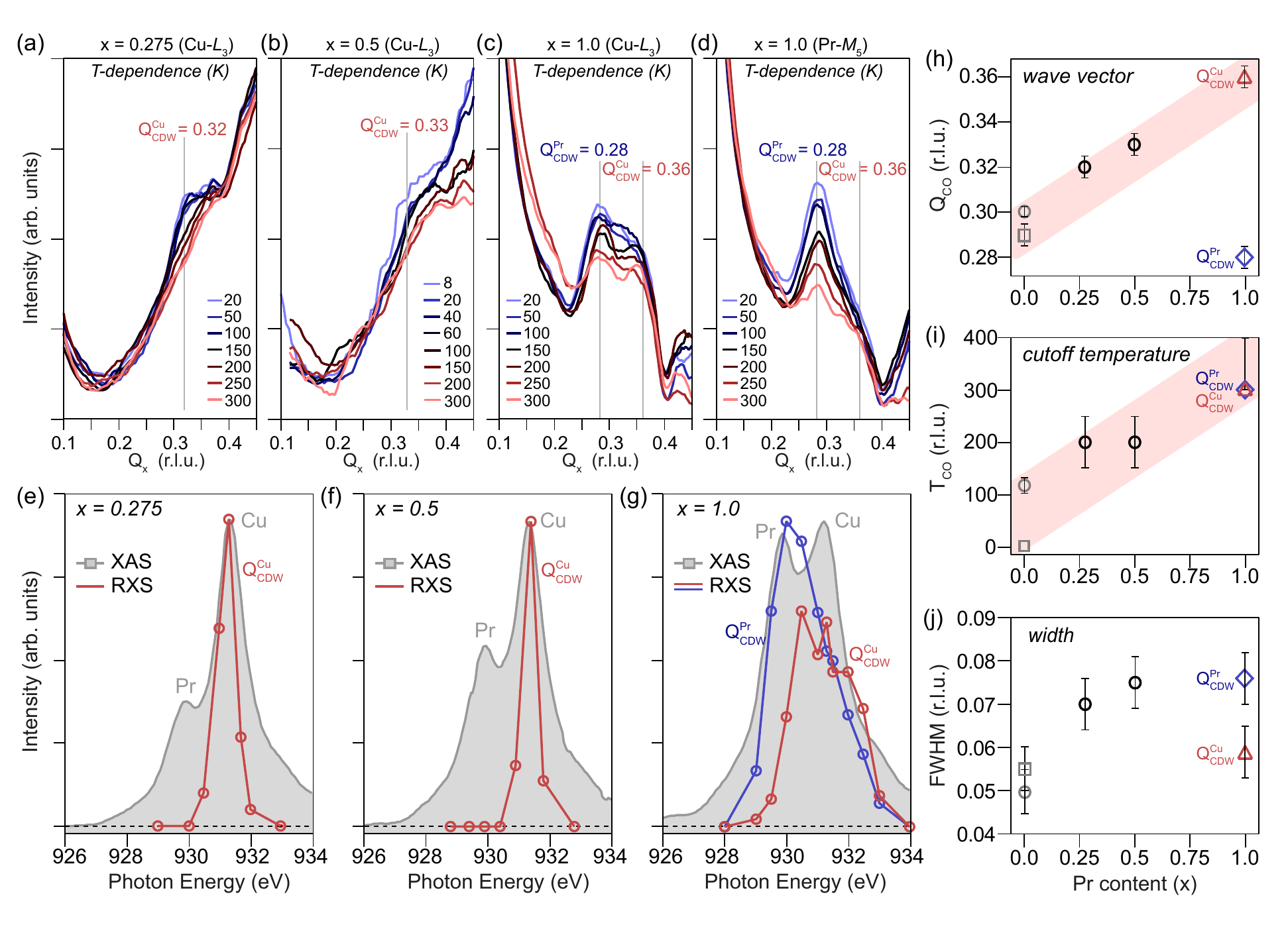}
\caption{\label{fig:3} Evolution of charge order in the Y$_{1-x}$Pr$_{x}$BCO series. (a,b) Temperature-dependent momentum scans from $x$ = 0.275 and $x$ = 0.5 Y$_{1-x}$Pr$_{x}$BCO films, measured at Cu-$L_{3}$ edge. (c,d) Temperature-dependent momentum scans from PrBCO films, measured at Cu-$L_{3}$ (c) and Pr-$M_{5}$ edge (d). (e-g), Resonant profile of the charge order peaks in $x$ = 0.275 (e), $x$ = 0.5 (f), and $x$ = 1 (g). In (g), the resonant behavior of $Q_{\mathrm{CDW}}^{\mathrm{Cu}}$ and $Q_{\mathrm{CDW}}^{\mathrm{Pr}}$ is displayed by the red and blue lines. Overlaid filled grey curves represent the X-ray absorption spectra of each sample. (h-j) Doping-dependence of wave vectors (h), cutoff temperatures (i), and widths (j) of the charge order peaks in Y$_{1-x}$Pr$_{x}$BCO. YBCO ($x$ = 0, $\delta \approx$ 0) data are adopted from previous RXS studies (Ref. \cite{Blanco-Canosa2014,Frano2016}). The data points plotted with grey square and grey circle correspond to bulk YBCO (Ref. \cite{Blanco-Canosa2014}) and YBCO thin film (Ref. \cite{Frano2016}, 50 nm thick on La$_{2/3}$Ca$_{1/3}$MnO$_{3}$), respectively. Rose shades in (h,i) are guides-to-the-eye.}
\end{figure*}

\section{\label{sec:level1} IV. Photon energy and temperature dependence of the diffraction data}

The resonant profiles (X-ray scattering intensity vs. photon energy) for Y$_{1-x}$Pr$_{x}$BCO samples at intermediate doping levels are shown in Fig. 5(e) and 5(f). The presence of a single resonance (at the Cu-edge) in these samples, in combination with the linear dependence of the ordering vector on doping, lends further support to the close connection between the charge crystal $Q_{\mathrm{CDW}}^{\mathrm{Cu}}$ peak observed in PrBCO and the charge-density-wave signatures observed at higher hole doping.


For the temperature dependence, we first focus on PrBCO, where the diffraction signatures at the Cu and Pr edge display a differing temperature dependence. As shown in Fig. 5(c,d), the resonant scattering intensity $I_{\mathrm{CDW}}^{\mathrm{Pr}}$ clearly decreases with increasing temperature, while $I_{\mathrm{CDW}}^{\mathrm{Cu}}$ evolves more gradually all the way up to 300 K (see Fig. S3 for details on the quantitative analysis of the temperature dependence \cite{SI}). This difference suggests that, while coexisting, these two modulated states are independent as it can be expected from their association to different degrees of freedom \cite{Kivelson1998,Vojta1999,Sachdev2013,Davis2013,Lee2014,Wang2014,DallaTorre2015,Duong2017}.
We also note that the very gradual evolution of $I_{\mathrm{CDW}}^{\mathrm{Cu}}$ in PrBCO is reminiscent of the temperature dependence of charge density fluctuations seen in \cite{arpaia_dynamical_2019} and therefore could suggest a partly fluctuating nature of the observed spatial modulations. The temperature dependence for the intermediate doping levels shows an evolution of $I_{\mathrm{CDW}}^{\mathrm{Cu}}$ consistent with a very gradual onset of the CDW signal around 150-200\,K, as observed in bulk YBCO in previous studies \cite{Blanco-Canosa2014}.

To summarize the trends for the cutoff temperature and the correlation lengths, while the former tends to increase with Pr-doping [Fig. 5(i)], the latter remains similarly short-ranged for all doping levels [Fig. 5(j)].

\section{\label{sec:level1} V. Charge order in the magnetic insulating limit}

In Figure 6, we summarize our results in the extended phase diagram of \textit{R}BCO. The latter includes long-range Cu (Neel) spin order (from our $\mu$SR data and Ref. \cite{Radousky1992,Akhavan2002}), low-temperature Pr spin order (from Ref. \cite{Radousky1992,Akhavan2002,Hill2000}), superconductivity [from our transport data in Fig. 1(e) and Ref. \cite{Radousky1992,Akhavan2002}], pseudogap [from resistivity anomalies, Fig. 1(e)], and charge order (from our RXS data, Figs. 2-5). 

The present \textit{R}BCO phase diagram demonstrates the coexistence between static Cu charge and spin orders in the very lightly-doped limit. Combined with the recent observation of electronic modulation on a very lightly-doped insulating surface of Bi$_2$Sr$_2$CaCu$_{2}$O$_{8+y}$ \cite{zhao_charge-stripe_2019}, our results support the existence of a common ’charge crystal’ phase emerging in the strongly-correlated ground state of lightly-doped CuO 2 planes. This observation is highly significant for our understanding of electronic symmetry-breaking in cuprates, as it reveals that electronic liquid or crystal phases pervade the phase diagram of the CuO$_{2}$ planes all the way into the Mott limit.

\begin{figure}
\includegraphics[width=0.95 \columnwidth]{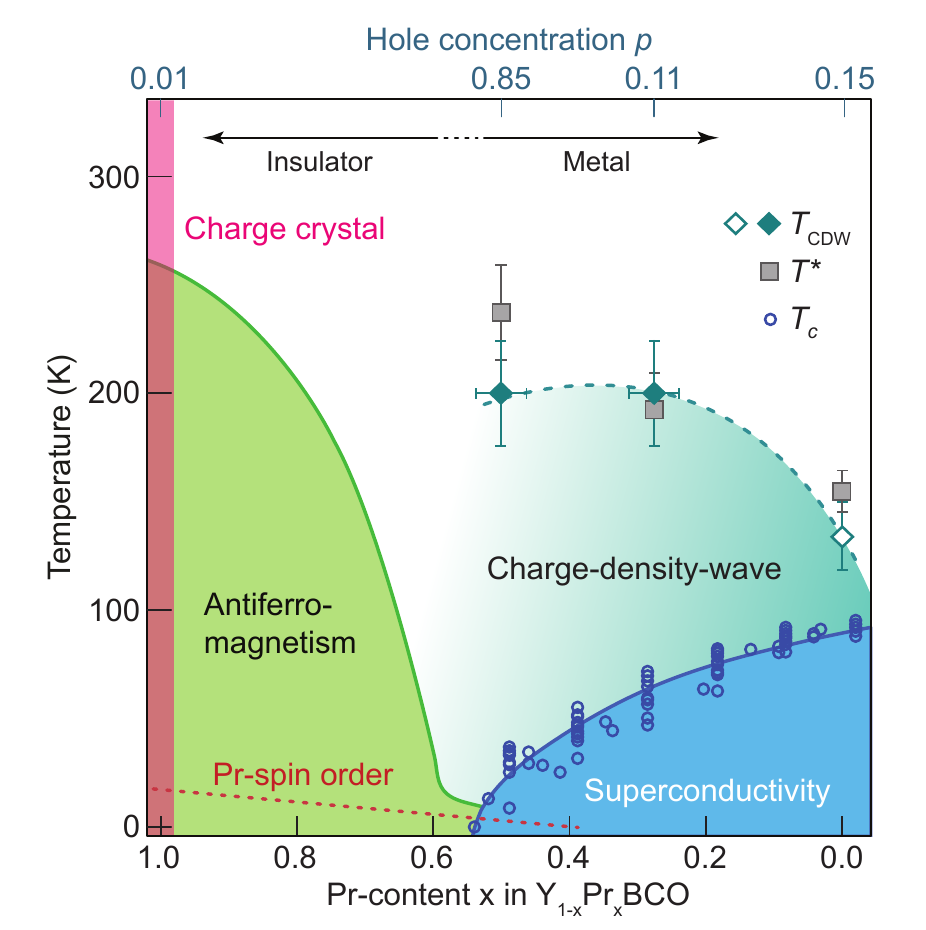}
\caption{\label{fig:epsart} Extended phase diagram of \textit{R}BCO. Green region denotes long-range Cu (Neel) spin order (from our $\mu$SR data and Ref. \cite{Radousky1992,Akhavan2002}). Dashed red line marks the onset of low-temperature Pr spin order (from Ref. \cite{Radousky1992,Akhavan2002,Hill2000}). Blue region denotes superconductivity (from transport data in Fig. 1(e) and Ref. \cite{Radousky1992,Akhavan2002}). Cyan region denotes charge order (from RXS data, Figs. 2-5). Magenta line on the left edge denotes the “charge crystal” phase. Grey squares mark the pseudogap onset temperature (from non-linear deviations in resistivity, see Fig. S5 in supplementary material). The top horizontal axis (hole concentration $p$) uses the $p$ values as extrapolated from the doping-dependence of the CDW wavevector $Q_{\mathrm{CDW}}$ (see text and Fig. 4). The ${T}_{\mathrm{CDW}}$ data point at $x=0$ and $p=0.15$ is from Ref. \cite{Blanco-Canosa2014} and is consistent with our results on HoBCO and EuBCO (see Supplementary Figure 4).}
\end{figure}


Regarding the origin of charge order in cuprates, two main viewpoints have been adopted over the years \cite{Kivelson1998,Vojta1999,Sachdev2013,Davis2013,Lee2014,Wang2014,DallaTorre2015,Duong2017}: the real space Coulomb-frustration scenario; and the momentum-space instability scenario. These competing hypotheses have been invoked to successfully explain different aspects of the phenomenology of charge order in cuprates. In this context, the present results offer important new insights, as the observation of a spatially modulated electronic phase in the insulating limit is fundamentally incompatible with the momentum-space instability scenario due to the absence of low-energy itinerant states in a gapped system. On the other hand, the real-space Coulomb-frustration scenario has the merit that it does not require the presence of a Fermi surface and is compatible with the insulating ground state of PrBCO.

\section{\label{sec:level1} Acknowledgements}

This work was supported by the Air Force Office of Scientific Research Young Investigator Program under grant FA9550-19-1-0063 and National Research Foundation of Korea, Ministry of Science, Grant No. 2016K1A4A4A01922028. Work at the University of Toronto was supported by the Natural Sciences and Engineering Research Council (NSERC) and the Canada Foundation for Innovation (CFI).  Part of the research described in this paper was performed at the Canadian Light Source, a national research facility of the University of Saskatchewan, which is supported by CFI, NSERC, the National Research Council (NRC), the Canadian Institutes of Health Research (CIHR), the Government of Saskatchewan, and the University of Saskatchewan. M.K. acknowledges a Samsung Scholarship from the Samsung Foundation of Culture

\section{\label{sec:level1} Appendix: Materials and methods}

\subsection{\label{sec:level2} 1. Pulsed Laser Deposition of YPrBCO thin films}

(Y,Pr)BCO thin films used in this study were epitaxially grown along the $c$-axis on (LaAlO$_{3}$)$_{0.3}$(Sr$_{2}$TaAlO$_{6}$)$_{0.7}$ (LSAT) substrates using pulsed laser-ablated deposition (PLD). LSAT was chosen as substrate material for this study because its in-plane lattice constant (3.868 \AA) is well matched with the in-plane lattice constants of (Y,Pr)BCO films \cite{Zhang2013}. The PLD growth was done using a 248 nm KrF excimer laser with a fluence of 2 J/cm$^{2}$. Films were grown using a substrate temperature of 800 °C and 200 mTorr O$_{2}$ chamber pressure. Further details of the PLD growth process are given in Ref. \cite{Zhang2018}. All the films grown had nominal thickness of 40 nm, which was set using target-specific growth rates. The growth rates were determined using atomic force microscopy, by measuring the step heights of test films after they were chemically etched to create a step edge. The films are expected to be $\approx$ 50 \% twinned, following the results of other similarly grown YBCO films (see Ref. \cite{Paturi2004}).

\subsection{\label{sec:level2} 2. Plasma assisted reactive sputter deposition of HoBCO and EuBCO films}

HoBCO  and EuBCO films were deposited on LSAT (100), polished single crystal substrates. The substrates were heated from 825 °C during deposition. A DC magnetron sputtering source was used with 50 mm HoBCO and EuBCO stoichiometric targets that had been verified by XRD. The chamber during growth was 330 mTorr with an Argon to Oxygen ratio of 1.5 sccm. Once the chamber settings were stable the DC magnetron sputter gun was ignited and ramped up to 55 watts over the course of 10 minutes. Typical growth rates were around 1 nm/min with a final film thickness of 55 nm. After deposition the chamber was filled with 500 Torr of ultra-high purity oxygen and allowed to stand at deposition temperatures for 10 minutes. After, power was cut from the heater and the samples were allowed to cool to room temperature in 500 Torr of oxygen. The critical temperatures were 92-94K with current density of 3MA/cm$^{2}$ at 77 K. Electrical transport measurements for HoBaCuO Josephson devices are reported here \cite{McCoy2019}.

\subsection{\label{sec:level2} 2. X-ray diffraction and transport characterization of YPrBCO films}

The as-grown films were characterized using X-ray diffraction (XRD) and resistance vs. temperature ($R$ vs. $T$) measurements.  XRD measurements were done using the $\theta$-2$\theta$ method and show the expected substrate and film peaks. The $R$ vs. $T$ measurements were done in a He-4 cryostat using an AC resistance bridge and silver contacts in the Van der Pauw configuration. Superconducting transitions were observed in the $x$ = 0.275 and $x$ = 0.5 films, and insulating behavior in the $x$ = 1 film. The critical temperatures ($T_{c}$) and $R$ vs. $T$ behaviors are similar to other YPrBCO films of comparable Pr and oxygen doping (see Refs. \cite{Radousky1992,Vovk2013,Antognazza1990,Peng1989}). Similarly grown YBCO films were shown to be epitaxially bonded to the substrate using transmission electron microscopy (TEM) in a previous study \cite{Zhang2013}.

\subsection{\label{sec:level2} 3. Resonant X-ray scattering experiments}

Resonant X-ray scattering experiments RXS experiments were performed in two different beamlines, the UE46-PGM1 beamline of BESSY II at Helmholtz Zentrum Berlin ($x$ = 0.275 and 1) and the REIXS (10ID-2) beamline of Canadian Light Source ($x$ = 0.5). All RXS measurements were conducted in ultra-high vacuum diffractometers with the pressure better than 10$^{-9}$ torr. Samples were aligned $in situ$ using (0 0 2) and (1 0 2) Bragg diffractions. Momentum-space scans were obtained by rocking the sample angle at a fixed detector position. X-ray absorption spectra (XAS) were measured by monitoring total electron yields. We used out-of-scattering-plane ($\sigma$) polarizations for both RXS and XAS measurements unless specified.

\subsection{\label{sec:level2} 4. Muon spin spectroscopy experiments}

The Low Energy $\mu$SR experiments where performed at the $\mu$E4 beamline \cite{muE4}
which is part of the Swiss Muon Source, Paul Scherrer Institut, Villigen, Switzerland.
The $\mu$SR data have been analyzed with the help of MUSRFIT \cite{musrfit}. Details about
the $\mu$SR technique can be found in Ref.\,\cite{yaouanc}.
All the measurements were conducted in ultra-high vacuum at a pressure smaller than $10^{-8}$ mbar
and zero magnetic field conditions. The decay positron asymmetry spectra were analyzed
with following function
$$ A(t) = A_{\rm osc}\, \cos(\gamma_\mu B_{\rm int} t + \phi) \cdot e^{-\lambda_{\rm T} t} + A_1\, e^{-(\lambda_{\rm L} t)^\beta}, $$
where $A_{\rm osc}$ is the coherent zero field oscillation amplitude, $\gamma_\mu$ is the muon gyromagnetic ratio, $B_{\rm int}$ the internal magnetic field at the muon site, and $\phi$ is the phase between the initial muon spin direction and the decay positron detector. $\lambda_{\rm T,L}$ are the transverse and longitudinal depolarization rates. $\beta \simeq 1$ except close to the phase transition where it is close to 2.



\begin{thebibliography}{52}%
\makeatletter
\providecommand \@ifxundefined [1]{%
 \@ifx{#1\undefined}
}%
\providecommand \@ifnum [1]{%
 \ifnum #1\expandafter \@firstoftwo
 \else \expandafter \@secondoftwo
 \fi
}%
\providecommand \@ifx [1]{%
 \ifx #1\expandafter \@firstoftwo
 \else \expandafter \@secondoftwo
 \fi
}%
\providecommand \natexlab [1]{#1}%
\providecommand \enquote  [1]{``#1''}%
\providecommand \bibnamefont  [1]{#1}%
\providecommand \bibfnamefont [1]{#1}%
\providecommand \citenamefont [1]{#1}%
\providecommand \href@noop [0]{\@secondoftwo}%
\providecommand \href [0]{\begingroup \@sanitize@url \@href}%
\providecommand \@href[1]{\@@startlink{#1}\@@href}%
\providecommand \@@href[1]{\endgroup#1\@@endlink}%
\providecommand \@sanitize@url [0]{\catcode `\\12\catcode `\$12\catcode
  `\&12\catcode `\#12\catcode `\^12\catcode `\_12\catcode `\%12\relax}%
\providecommand \@@startlink[1]{}%
\providecommand \@@endlink[0]{}%
\providecommand \url  [0]{\begingroup\@sanitize@url \@url }%
\providecommand \@url [1]{\endgroup\@href {#1}{\urlprefix }}%
\providecommand \urlprefix  [0]{URL }%
\providecommand \Eprint [0]{\href }%
\providecommand \doibase [0]{https://doi.org/}%
\providecommand \selectlanguage [0]{\@gobble}%
\providecommand \bibinfo  [0]{\@secondoftwo}%
\providecommand \bibfield  [0]{\@secondoftwo}%
\providecommand \translation [1]{[#1]}%
\providecommand \BibitemOpen [0]{}%
\providecommand \bibitemStop [0]{}%
\providecommand \bibitemNoStop [0]{.\EOS\space}%
\providecommand \EOS [0]{\spacefactor3000\relax}%
\providecommand \BibitemShut  [1]{\csname bibitem#1\endcsname}%
\let\auto@bib@innerbib\@empty
\bibitem [{\citenamefont {Bednorz}\ and\ \citenamefont
  {MÃ¼ller}(1986)}]{bednorz_possible_1986}%
  \BibitemOpen
  \bibfield  {author} {\bibinfo {author} {\bibfnamefont {J.~G.}\ \bibnamefont
  {Bednorz}}\ and\ \bibinfo {author} {\bibfnamefont {K.~A.}\ \bibnamefont
  {MÃ¼ller}},\ }\bibfield  {title} {\bibinfo {title} {Possible {highT} c
  superconductivity in the {Ba}-{La}-{Cu}-{O} system},\ }\href
  {https://doi.org/10.1007/BF01303701} {\bibfield  {journal} {\bibinfo
  {journal} {Zeitschrift fÃ¼r Physik B Condensed Matter}\ }\textbf {\bibinfo
  {volume} {64}},\ \bibinfo {pages} {189} (\bibinfo {year} {1986})}\BibitemShut
  {NoStop}%
\bibitem [{\citenamefont {Zaanen}\ and\ \citenamefont
  {Gunnarsson}(1989)}]{zaanen_charged_1989}%
  \BibitemOpen
  \bibfield  {author} {\bibinfo {author} {\bibfnamefont {J.}~\bibnamefont
  {Zaanen}}\ and\ \bibinfo {author} {\bibfnamefont {O.}~\bibnamefont
  {Gunnarsson}},\ }\bibfield  {title} {\bibinfo {title} {Charged magnetic
  domain lines and the magnetism of high-{Tc} oxides},\ }\href@noop {}
  {\bibfield  {journal} {\bibinfo  {journal} {Phys. Rev. B}\ }\textbf {\bibinfo
  {volume} {40}},\ \bibinfo {pages} {7391} (\bibinfo {year}
  {1989})}\BibitemShut {NoStop}%
\bibitem [{\citenamefont {Machida}(1989)}]{machida_magnetism_1989}%
  \BibitemOpen
  \bibfield  {author} {\bibinfo {author} {\bibfnamefont {K.}~\bibnamefont
  {Machida}},\ }\bibfield  {title} {\bibinfo {title} {Magnetism in {La2CuO4}
  based compounds},\ }\href@noop {} {\bibfield  {journal} {\bibinfo  {journal}
  {Physica C: Superconductivity}\ }\textbf {\bibinfo {volume} {158}},\ \bibinfo
  {pages} {192 } (\bibinfo {year} {1989})}\BibitemShut {NoStop}%
\bibitem [{\citenamefont {Poilblanc}\ and\ \citenamefont
  {Rice}(1989)}]{poilblanc_charged_1989}%
  \BibitemOpen
  \bibfield  {author} {\bibinfo {author} {\bibfnamefont {D.}~\bibnamefont
  {Poilblanc}}\ and\ \bibinfo {author} {\bibfnamefont {T.~M.}\ \bibnamefont
  {Rice}},\ }\bibfield  {title} {\bibinfo {title} {Charged solitons in the
  {Hartree}-{Fock} approximation to the large- {\textbackslash}textit\{{U}\}
  {Hubbard} model},\ }\href {https://doi.org/10.1103/PhysRevB.39.9749}
  {\bibfield  {journal} {\bibinfo  {journal} {Physical Review B}\ }\textbf
  {\bibinfo {volume} {39}},\ \bibinfo {pages} {9749} (\bibinfo {year}
  {1989})}\BibitemShut {NoStop}%
\bibitem [{\citenamefont {Emery}\ \emph {et~al.}(1990)\citenamefont {Emery},
  \citenamefont {Kivelson},\ and\ \citenamefont {Lin}}]{emery_phase_1990}%
  \BibitemOpen
  \bibfield  {author} {\bibinfo {author} {\bibfnamefont {V.~J.}\ \bibnamefont
  {Emery}}, \bibinfo {author} {\bibfnamefont {S.~A.}\ \bibnamefont
  {Kivelson}},\ and\ \bibinfo {author} {\bibfnamefont {H.~Q.}\ \bibnamefont
  {Lin}},\ }\bibfield  {title} {\bibinfo {title} {Phase separation in the t-{J}
  model},\ }\href {https://doi.org/10.1103/PhysRevLett.64.475} {\bibfield
  {journal} {\bibinfo  {journal} {Physical Review Letters}\ }\textbf {\bibinfo
  {volume} {64}},\ \bibinfo {pages} {475} (\bibinfo {year} {1990})}\BibitemShut
  {NoStop}%
\bibitem [{\citenamefont {Tranquada}\ \emph {et~al.}(1995)\citenamefont
  {Tranquada}, \citenamefont {Sternlieb}, \citenamefont {Axe}, \citenamefont
  {Nakamura},\ and\ \citenamefont {Uchida}}]{tranquada_stripes_1995}%
  \BibitemOpen
  \bibfield  {author} {\bibinfo {author} {\bibfnamefont {J.~M.}\ \bibnamefont
  {Tranquada}}, \bibinfo {author} {\bibfnamefont {B.~J.}\ \bibnamefont
  {Sternlieb}}, \bibinfo {author} {\bibfnamefont {J.~D.}\ \bibnamefont {Axe}},
  \bibinfo {author} {\bibfnamefont {Y.}~\bibnamefont {Nakamura}},\ and\
  \bibinfo {author} {\bibfnamefont {S.}~\bibnamefont {Uchida}},\ }\bibfield
  {title} {\bibinfo {title} {Evidence for stripe correlations of spins and
  holes in copper oxide superconductors},\ }\href
  {https://doi.org/10.1038/375561a0} {\bibfield  {journal} {\bibinfo  {journal}
  {Nature}\ }\textbf {\bibinfo {volume} {375}},\ \bibinfo {pages} {561}
  (\bibinfo {year} {1995})}\BibitemShut {NoStop}%
\bibitem [{\citenamefont {Ghiringhelli}\ \emph {et~al.}(2012)\citenamefont
  {Ghiringhelli}, \citenamefont {Tacon}, \citenamefont {Minola}, \citenamefont
  {Mazzoli}, \citenamefont {Brookes}, \citenamefont {Luca}, \citenamefont
  {Frano}, \citenamefont {Hawthorn}, \citenamefont {He}, \citenamefont {Loew},
  \citenamefont {Sala}, \citenamefont {Peets}, \citenamefont {Salluzzo},
  \citenamefont {Schierle}, \citenamefont {Sutarto}, \citenamefont {Sawatzky},
  \citenamefont {Weschke}, \citenamefont {Keimer},\ and\ \citenamefont
  {Braicovich}}]{Ghiringhelli2012}%
  \BibitemOpen
  \bibfield  {author} {\bibinfo {author} {\bibfnamefont {G.}~\bibnamefont
  {Ghiringhelli}}, \bibinfo {author} {\bibfnamefont {M.~L.}\ \bibnamefont
  {Tacon}}, \bibinfo {author} {\bibfnamefont {M.}~\bibnamefont {Minola}},
  \bibinfo {author} {\bibfnamefont {C.}~\bibnamefont {Mazzoli}}, \bibinfo
  {author} {\bibfnamefont {N.~B.}\ \bibnamefont {Brookes}}, \bibinfo {author}
  {\bibfnamefont {G.~M.~D.}\ \bibnamefont {Luca}}, \bibinfo {author}
  {\bibfnamefont {A.}~\bibnamefont {Frano}}, \bibinfo {author} {\bibfnamefont
  {D.~G.}\ \bibnamefont {Hawthorn}}, \bibinfo {author} {\bibfnamefont
  {F.}~\bibnamefont {He}}, \bibinfo {author} {\bibfnamefont {T.}~\bibnamefont
  {Loew}}, \bibinfo {author} {\bibfnamefont {M.~M.}\ \bibnamefont {Sala}},
  \bibinfo {author} {\bibfnamefont {D.~C.}\ \bibnamefont {Peets}}, \bibinfo
  {author} {\bibfnamefont {M.}~\bibnamefont {Salluzzo}}, \bibinfo {author}
  {\bibfnamefont {E.}~\bibnamefont {Schierle}}, \bibinfo {author}
  {\bibfnamefont {R.}~\bibnamefont {Sutarto}}, \bibinfo {author} {\bibfnamefont
  {G.~a.}\ \bibnamefont {Sawatzky}}, \bibinfo {author} {\bibfnamefont
  {E.}~\bibnamefont {Weschke}}, \bibinfo {author} {\bibfnamefont
  {B.}~\bibnamefont {Keimer}},\ and\ \bibinfo {author} {\bibfnamefont
  {L.}~\bibnamefont {Braicovich}},\ }\bibfield  {title} {\bibinfo {title}
  {{Long-Range Incommensurate Charge Fluctuations in (Y,Nd)Ba2Cu3O6+x}},\
  }\href@noop {} {\bibfield  {journal} {\bibinfo  {journal} {Science}\ }\textbf
  {\bibinfo {volume} {337}},\ \bibinfo {pages} {821} (\bibinfo {year}
  {2012})}\BibitemShut {NoStop}%
\bibitem [{\citenamefont {Chang}\ \emph {et~al.}(2012)\citenamefont {Chang},
  \citenamefont {Blackburn}, \citenamefont {Holmes}, \citenamefont
  {Christensen}, \citenamefont {Larsen}, \citenamefont {Mesot}, \citenamefont
  {Liang}, \citenamefont {Bonn}, \citenamefont {Hardy}, \citenamefont
  {Watenphul}, \citenamefont {Zimmermann}, \citenamefont {Forgan},\ and\
  \citenamefont {Hayden}}]{Chang2012}%
  \BibitemOpen
  \bibfield  {author} {\bibinfo {author} {\bibfnamefont {J.}~\bibnamefont
  {Chang}}, \bibinfo {author} {\bibfnamefont {E.}~\bibnamefont {Blackburn}},
  \bibinfo {author} {\bibfnamefont {A.~T.}\ \bibnamefont {Holmes}}, \bibinfo
  {author} {\bibfnamefont {N.~B.}\ \bibnamefont {Christensen}}, \bibinfo
  {author} {\bibfnamefont {J.}~\bibnamefont {Larsen}}, \bibinfo {author}
  {\bibfnamefont {J.}~\bibnamefont {Mesot}}, \bibinfo {author} {\bibfnamefont
  {R.}~\bibnamefont {Liang}}, \bibinfo {author} {\bibfnamefont {D.~a.}\
  \bibnamefont {Bonn}}, \bibinfo {author} {\bibfnamefont {W.~N.}\ \bibnamefont
  {Hardy}}, \bibinfo {author} {\bibfnamefont {A.}~\bibnamefont {Watenphul}},
  \bibinfo {author} {\bibfnamefont {M.~V.}\ \bibnamefont {Zimmermann}},
  \bibinfo {author} {\bibfnamefont {E.~M.}\ \bibnamefont {Forgan}},\ and\
  \bibinfo {author} {\bibfnamefont {S.~M.}\ \bibnamefont {Hayden}},\ }\bibfield
   {title} {\bibinfo {title} {{Direct observation of competition between
  superconductivity and charge density wave order in YBa2Cu3O6.67}},\ }\href
  {https://doi.org/10.1038/nphys2456} {\bibfield  {journal} {\bibinfo
  {journal} {Nature Physics}\ }\textbf {\bibinfo {volume} {8}},\ \bibinfo
  {pages} {871} (\bibinfo {year} {2012})},\ \Eprint
  {https://arxiv.org/abs/1206.4333} {arXiv:1206.4333} \BibitemShut {NoStop}%
\bibitem [{\citenamefont {Blanco-Canosa}\ \emph {et~al.}(2014)\citenamefont
  {Blanco-Canosa}, \citenamefont {Frano}, \citenamefont {Schierle},
  \citenamefont {Porras}, \citenamefont {Loew}, \citenamefont {Minola},
  \citenamefont {Bluschke}, \citenamefont {Weschke}, \citenamefont {Keimer},\
  and\ \citenamefont {Tacon}}]{Blanco-Canosa2014}%
  \BibitemOpen
  \bibfield  {author} {\bibinfo {author} {\bibfnamefont {S.}~\bibnamefont
  {Blanco-Canosa}}, \bibinfo {author} {\bibfnamefont {A.}~\bibnamefont
  {Frano}}, \bibinfo {author} {\bibfnamefont {E.}~\bibnamefont {Schierle}},
  \bibinfo {author} {\bibfnamefont {J.}~\bibnamefont {Porras}}, \bibinfo
  {author} {\bibfnamefont {T.}~\bibnamefont {Loew}}, \bibinfo {author}
  {\bibfnamefont {M.}~\bibnamefont {Minola}}, \bibinfo {author} {\bibfnamefont
  {M.}~\bibnamefont {Bluschke}}, \bibinfo {author} {\bibfnamefont
  {E.}~\bibnamefont {Weschke}}, \bibinfo {author} {\bibfnamefont
  {B.}~\bibnamefont {Keimer}},\ and\ \bibinfo {author} {\bibfnamefont {M.~L.}\
  \bibnamefont {Tacon}},\ }\bibfield  {title} {\bibinfo {title} {{Resonant
  X-ray Scattering Study of Charge Density Wave Correlations in YBa2Cu3O6+x}},\
  }\href {https://doi.org/10.1103/PhysRevB.90.054513} {\bibfield  {journal}
  {\bibinfo  {journal} {Physical Review B}\ }\textbf {\bibinfo {volume} {90}},\
  \bibinfo {pages} {054513} (\bibinfo {year} {2014})},\ \Eprint
  {https://arxiv.org/abs/1406.1595} {arXiv:1406.1595} \BibitemShut {NoStop}%
\bibitem [{\citenamefont {Comin}\ \emph {et~al.}(2014)\citenamefont {Comin},
  \citenamefont {Frano}, \citenamefont {Yee}, \citenamefont {Yoshida},
  \citenamefont {Eisaki}, \citenamefont {Schierle}, \citenamefont {Weschke},
  \citenamefont {Sutarto}, \citenamefont {He}, \citenamefont {Soumyanarayanan},
  \citenamefont {He}, \citenamefont {{Le Tacon}}, \citenamefont {Elfimov},
  \citenamefont {Hoffman}, \citenamefont {Sawatzky}, \citenamefont {Keimer},\
  and\ \citenamefont {Damascelli}}]{Comin2014}%
  \BibitemOpen
  \bibfield  {author} {\bibinfo {author} {\bibfnamefont {R.}~\bibnamefont
  {Comin}}, \bibinfo {author} {\bibfnamefont {A.}~\bibnamefont {Frano}},
  \bibinfo {author} {\bibfnamefont {M.~M.}\ \bibnamefont {Yee}}, \bibinfo
  {author} {\bibfnamefont {Y.}~\bibnamefont {Yoshida}}, \bibinfo {author}
  {\bibfnamefont {H.}~\bibnamefont {Eisaki}}, \bibinfo {author} {\bibfnamefont
  {E.}~\bibnamefont {Schierle}}, \bibinfo {author} {\bibfnamefont
  {E.}~\bibnamefont {Weschke}}, \bibinfo {author} {\bibfnamefont
  {R.}~\bibnamefont {Sutarto}}, \bibinfo {author} {\bibfnamefont
  {F.}~\bibnamefont {He}}, \bibinfo {author} {\bibfnamefont {A.}~\bibnamefont
  {Soumyanarayanan}}, \bibinfo {author} {\bibfnamefont {Y.}~\bibnamefont {He}},
  \bibinfo {author} {\bibfnamefont {M.}~\bibnamefont {{Le Tacon}}}, \bibinfo
  {author} {\bibfnamefont {I.~S.}\ \bibnamefont {Elfimov}}, \bibinfo {author}
  {\bibfnamefont {J.~E.}\ \bibnamefont {Hoffman}}, \bibinfo {author}
  {\bibfnamefont {G.~A.}\ \bibnamefont {Sawatzky}}, \bibinfo {author}
  {\bibfnamefont {B.}~\bibnamefont {Keimer}},\ and\ \bibinfo {author}
  {\bibfnamefont {A.}~\bibnamefont {Damascelli}},\ }\bibfield  {title}
  {\bibinfo {title} {{Charge Order Driven by Fermi-Arc Instability in
  Bi2Sr2-xLaxCuO6+d}},\ }\href@noop {} {\bibfield  {journal} {\bibinfo
  {journal} {Science}\ }\textbf {\bibinfo {volume} {343}},\ \bibinfo {pages}
  {390} (\bibinfo {year} {2014})}\BibitemShut {NoStop}%
\bibitem [{\citenamefont {{da Silva Neto}}\ \emph {et~al.}(2014)\citenamefont
  {{da Silva Neto}}, \citenamefont {Aynajian}, \citenamefont {Frano},
  \citenamefont {Comin}, \citenamefont {Schierle}, \citenamefont {Weschke},
  \citenamefont {Gyenis}, \citenamefont {Jinsheng}, \citenamefont {Schneeloch},
  \citenamefont {Xu}, \citenamefont {Ono}, \citenamefont {Gu}, \citenamefont
  {{Le Tacon}},\ and\ \citenamefont {Yazdani}}]{DaSilvaNeto2014}%
  \BibitemOpen
  \bibfield  {author} {\bibinfo {author} {\bibfnamefont {E.~H.}\ \bibnamefont
  {{da Silva Neto}}}, \bibinfo {author} {\bibfnamefont {P.}~\bibnamefont
  {Aynajian}}, \bibinfo {author} {\bibfnamefont {A.}~\bibnamefont {Frano}},
  \bibinfo {author} {\bibfnamefont {R.}~\bibnamefont {Comin}}, \bibinfo
  {author} {\bibfnamefont {E.}~\bibnamefont {Schierle}}, \bibinfo {author}
  {\bibfnamefont {E.}~\bibnamefont {Weschke}}, \bibinfo {author} {\bibfnamefont
  {A.}~\bibnamefont {Gyenis}}, \bibinfo {author} {\bibfnamefont
  {W.}~\bibnamefont {Jinsheng}}, \bibinfo {author} {\bibfnamefont
  {J.}~\bibnamefont {Schneeloch}}, \bibinfo {author} {\bibfnamefont
  {Z.}~\bibnamefont {Xu}}, \bibinfo {author} {\bibfnamefont {S.}~\bibnamefont
  {Ono}}, \bibinfo {author} {\bibfnamefont {G.}~\bibnamefont {Gu}}, \bibinfo
  {author} {\bibfnamefont {M.}~\bibnamefont {{Le Tacon}}},\ and\ \bibinfo
  {author} {\bibfnamefont {A.}~\bibnamefont {Yazdani}},\ }\bibfield  {title}
  {\bibinfo {title} {{Ubiquitous interplay between charge ordering and high
  temperature superconductivity in cuprates}},\ }\href
  {https://doi.org/10.1038/ncomms1354} {\bibfield  {journal} {\bibinfo
  {journal} {Science (New York, N.Y.)}\ }\textbf {\bibinfo {volume} {343}},\
  \bibinfo {pages} {393} (\bibinfo {year} {2014})}\BibitemShut {NoStop}%
\bibitem [{\citenamefont {Tabis}\ \emph {et~al.}(2014)\citenamefont {Tabis},
  \citenamefont {Li}, \citenamefont {Tacon}, \citenamefont {Braicovich},
  \citenamefont {Kreyssig}, \citenamefont {Minola}, \citenamefont {Dellea},
  \citenamefont {Weschke}, \citenamefont {Veit}, \citenamefont {Ramazanoglu},
  \citenamefont {Goldman}, \citenamefont {Schmitt}, \citenamefont
  {Ghiringhelli}, \citenamefont {Bari{\v{s}}i{\'{c}}}, \citenamefont {Chan},
  \citenamefont {Dorow}, \citenamefont {Yu}, \citenamefont {Zhao},
  \citenamefont {Keimer},\ and\ \citenamefont {Greven}}]{Tabis2014}%
  \BibitemOpen
  \bibfield  {author} {\bibinfo {author} {\bibfnamefont {W.}~\bibnamefont
  {Tabis}}, \bibinfo {author} {\bibfnamefont {Y.}~\bibnamefont {Li}}, \bibinfo
  {author} {\bibfnamefont {M.~L.}\ \bibnamefont {Tacon}}, \bibinfo {author}
  {\bibfnamefont {L.}~\bibnamefont {Braicovich}}, \bibinfo {author}
  {\bibfnamefont {A.}~\bibnamefont {Kreyssig}}, \bibinfo {author}
  {\bibfnamefont {M.}~\bibnamefont {Minola}}, \bibinfo {author} {\bibfnamefont
  {G.}~\bibnamefont {Dellea}}, \bibinfo {author} {\bibfnamefont
  {E.}~\bibnamefont {Weschke}}, \bibinfo {author} {\bibfnamefont {M.~J.}\
  \bibnamefont {Veit}}, \bibinfo {author} {\bibfnamefont {M.}~\bibnamefont
  {Ramazanoglu}}, \bibinfo {author} {\bibfnamefont {A.~I.}\ \bibnamefont
  {Goldman}}, \bibinfo {author} {\bibfnamefont {T.}~\bibnamefont {Schmitt}},
  \bibinfo {author} {\bibfnamefont {G.}~\bibnamefont {Ghiringhelli}}, \bibinfo
  {author} {\bibfnamefont {N.}~\bibnamefont {Bari{\v{s}}i{\'{c}}}}, \bibinfo
  {author} {\bibfnamefont {M.~K.}\ \bibnamefont {Chan}}, \bibinfo {author}
  {\bibfnamefont {C.~J.}\ \bibnamefont {Dorow}}, \bibinfo {author}
  {\bibfnamefont {G.}~\bibnamefont {Yu}}, \bibinfo {author} {\bibfnamefont
  {X.}~\bibnamefont {Zhao}}, \bibinfo {author} {\bibfnamefont {B.}~\bibnamefont
  {Keimer}},\ and\ \bibinfo {author} {\bibfnamefont {M.}~\bibnamefont
  {Greven}},\ }\bibfield  {title} {\bibinfo {title} {{Charge order and its
  connection with Fermi-liquid charge transport in a pristine high-Tc
  cuprate}},\ }\href {https://doi.org/10.1038/ncomms6875} {\bibfield  {journal}
  {\bibinfo  {journal} {Nature Communications}\ }\textbf {\bibinfo {volume}
  {5}},\ \bibinfo {pages} {5875} (\bibinfo {year} {2014})},\ \Eprint
  {https://arxiv.org/abs/1404.7658} {arXiv:1404.7658} \BibitemShut {NoStop}%
\bibitem [{\citenamefont {da~silva Neto}\ \emph {et~al.}(2015)\citenamefont
  {da~silva Neto}, \citenamefont {Comin}, \citenamefont {He}, \citenamefont
  {Sutarto}, \citenamefont {Jiang}, \citenamefont {Greene}, \citenamefont
  {Sawatzky},\ and\ \citenamefont {Damascelli}}]{DasilvaNeto2015}%
  \BibitemOpen
  \bibfield  {author} {\bibinfo {author} {\bibfnamefont {E.~H.}\ \bibnamefont
  {da~silva Neto}}, \bibinfo {author} {\bibfnamefont {R.}~\bibnamefont
  {Comin}}, \bibinfo {author} {\bibfnamefont {F.}~\bibnamefont {He}}, \bibinfo
  {author} {\bibfnamefont {R.}~\bibnamefont {Sutarto}}, \bibinfo {author}
  {\bibfnamefont {Y.}~\bibnamefont {Jiang}}, \bibinfo {author} {\bibfnamefont
  {R.~L.}\ \bibnamefont {Greene}}, \bibinfo {author} {\bibfnamefont {G.~A.}\
  \bibnamefont {Sawatzky}},\ and\ \bibinfo {author} {\bibfnamefont
  {A.}~\bibnamefont {Damascelli}},\ }\bibfield  {title} {\bibinfo {title}
  {{Charge ordering in electron doped superconductor Nd2-xCexCuO4}},\
  }\href@noop {} {\bibfield  {journal} {\bibinfo  {journal} {Science}\ }\textbf
  {\bibinfo {volume} {347}},\ \bibinfo {pages} {282} (\bibinfo {year}
  {2015})}\BibitemShut {NoStop}%
\bibitem [{\citenamefont {{da Silva Neto}}\ \emph {et~al.}(2016)\citenamefont
  {{da Silva Neto}}, \citenamefont {Yu}, \citenamefont {Minola}, \citenamefont
  {Sutarto}, \citenamefont {Schierle}, \citenamefont {Boschini}, \citenamefont
  {Zonno}, \citenamefont {Bluschke}, \citenamefont {Higgins}, \citenamefont
  {Li}, \citenamefont {Yu}, \citenamefont {Weschke}, \citenamefont {He},
  \citenamefont {Tacon}, \citenamefont {Greene}, \citenamefont {Greven},
  \citenamefont {Sawatzky}, \citenamefont {Keimer},\ and\ \citenamefont
  {Damascelli}}]{DaSilvaNeto2016}%
  \BibitemOpen
  \bibfield  {author} {\bibinfo {author} {\bibfnamefont {E.~H.}\ \bibnamefont
  {{da Silva Neto}}}, \bibinfo {author} {\bibfnamefont {B.}~\bibnamefont {Yu}},
  \bibinfo {author} {\bibfnamefont {M.}~\bibnamefont {Minola}}, \bibinfo
  {author} {\bibfnamefont {R.}~\bibnamefont {Sutarto}}, \bibinfo {author}
  {\bibfnamefont {E.}~\bibnamefont {Schierle}}, \bibinfo {author}
  {\bibfnamefont {F.}~\bibnamefont {Boschini}}, \bibinfo {author}
  {\bibfnamefont {M.}~\bibnamefont {Zonno}}, \bibinfo {author} {\bibfnamefont
  {M.}~\bibnamefont {Bluschke}}, \bibinfo {author} {\bibfnamefont
  {J.}~\bibnamefont {Higgins}}, \bibinfo {author} {\bibfnamefont
  {Y.}~\bibnamefont {Li}}, \bibinfo {author} {\bibfnamefont {G.}~\bibnamefont
  {Yu}}, \bibinfo {author} {\bibfnamefont {E.}~\bibnamefont {Weschke}},
  \bibinfo {author} {\bibfnamefont {F.}~\bibnamefont {He}}, \bibinfo {author}
  {\bibfnamefont {M.~L.}\ \bibnamefont {Tacon}}, \bibinfo {author}
  {\bibfnamefont {R.~L.}\ \bibnamefont {Greene}}, \bibinfo {author}
  {\bibfnamefont {M.}~\bibnamefont {Greven}}, \bibinfo {author} {\bibfnamefont
  {G.~a.}\ \bibnamefont {Sawatzky}}, \bibinfo {author} {\bibfnamefont
  {B.}~\bibnamefont {Keimer}},\ and\ \bibinfo {author} {\bibfnamefont
  {A.}~\bibnamefont {Damascelli}},\ }\bibfield  {title} {\bibinfo {title}
  {{Doping dependent charge order correlations in electron-doped cuprates}},\
  }\href {https://doi.org/10.1126/sciadv.1600782} {\bibfield  {journal}
  {\bibinfo  {journal} {Science Advances}\ }\textbf {\bibinfo {volume} {2}},\
  \bibinfo {pages} {e1600782} (\bibinfo {year} {2016})},\ \Eprint
  {https://arxiv.org/abs/1607.06094} {arXiv:1607.06094} \BibitemShut {NoStop}%
\bibitem [{\citenamefont {Kang}\ \emph {et~al.}(2019)\citenamefont {Kang},
  \citenamefont {Pelliciari}, \citenamefont {Frano}, \citenamefont {Breznay},
  \citenamefont {Schierle}, \citenamefont {Weschke}, \citenamefont {Sutarto},
  \citenamefont {He}, \citenamefont {Shafer}, \citenamefont {Arenholz},
  \citenamefont {Chen}, \citenamefont {Zhang}, \citenamefont {Ruiz},
  \citenamefont {Hao}, \citenamefont {Lewin}, \citenamefont {Analytis},
  \citenamefont {Krockenberger}, \citenamefont {Yamamoto}, \citenamefont
  {Das},\ and\ \citenamefont {Comin}}]{Kang2019}%
  \BibitemOpen
  \bibfield  {author} {\bibinfo {author} {\bibfnamefont {M.}~\bibnamefont
  {Kang}}, \bibinfo {author} {\bibfnamefont {J.}~\bibnamefont {Pelliciari}},
  \bibinfo {author} {\bibfnamefont {A.}~\bibnamefont {Frano}}, \bibinfo
  {author} {\bibfnamefont {N.}~\bibnamefont {Breznay}}, \bibinfo {author}
  {\bibfnamefont {E.}~\bibnamefont {Schierle}}, \bibinfo {author}
  {\bibfnamefont {E.}~\bibnamefont {Weschke}}, \bibinfo {author} {\bibfnamefont
  {R.}~\bibnamefont {Sutarto}}, \bibinfo {author} {\bibfnamefont
  {F.}~\bibnamefont {He}}, \bibinfo {author} {\bibfnamefont {P.}~\bibnamefont
  {Shafer}}, \bibinfo {author} {\bibfnamefont {E.}~\bibnamefont {Arenholz}},
  \bibinfo {author} {\bibfnamefont {M.}~\bibnamefont {Chen}}, \bibinfo {author}
  {\bibfnamefont {K.}~\bibnamefont {Zhang}}, \bibinfo {author} {\bibfnamefont
  {A.}~\bibnamefont {Ruiz}}, \bibinfo {author} {\bibfnamefont {Z.}~\bibnamefont
  {Hao}}, \bibinfo {author} {\bibfnamefont {S.}~\bibnamefont {Lewin}}, \bibinfo
  {author} {\bibfnamefont {J.}~\bibnamefont {Analytis}}, \bibinfo {author}
  {\bibfnamefont {Y.}~\bibnamefont {Krockenberger}}, \bibinfo {author}
  {\bibfnamefont {H.}~\bibnamefont {Yamamoto}}, \bibinfo {author}
  {\bibfnamefont {T.}~\bibnamefont {Das}},\ and\ \bibinfo {author}
  {\bibfnamefont {R.}~\bibnamefont {Comin}},\ }\bibfield  {title} {\bibinfo
  {title} {{Evolution of charge order topology across a magnetic phase
  transition in cuprate superconductors}},\ }\href
  {https://doi.org/10.1038/s41567-018-0401-8} {\bibfield  {journal} {\bibinfo
  {journal} {Nature Physics}\ }\textbf {\bibinfo {volume} {15}},\ \bibinfo
  {pages} {335} (\bibinfo {year} {2019})}\BibitemShut {NoStop}%
\bibitem [{\citenamefont {Comin}(2016)}]{Comin2016}%
  \BibitemOpen
  \bibfield  {author} {\bibinfo {author} {\bibfnamefont {R.}~\bibnamefont
  {Comin}},\ }\bibfield  {title} {\bibinfo {title} {{Resonant X-ray scattering
  studies of charge order in cuprates}},\ }\href {https://doi.org/10.1146/))}
  {\bibfield  {journal} {\bibinfo  {journal} {Annu. Rev. Condens. Matter
  Phys.}\ }\textbf {\bibinfo {volume} {7}},\ \bibinfo {pages} {369} (\bibinfo
  {year} {2016})},\ \Eprint {https://arxiv.org/abs/1406.3533} {arXiv:1406.3533}
  \BibitemShut {NoStop}%
\bibitem [{\citenamefont {Kivelson}\ \emph {et~al.}(2003)\citenamefont
  {Kivelson}, \citenamefont {Bindloss}, \citenamefont {Fradkin}, \citenamefont
  {Oganesyan}, \citenamefont {Tranquada}, \citenamefont {Kapitulnik},\ and\
  \citenamefont {Howald}}]{kivelson_how_2003}%
  \BibitemOpen
  \bibfield  {author} {\bibinfo {author} {\bibfnamefont {S.~A.}\ \bibnamefont
  {Kivelson}}, \bibinfo {author} {\bibfnamefont {I.~P.}\ \bibnamefont
  {Bindloss}}, \bibinfo {author} {\bibfnamefont {E.}~\bibnamefont {Fradkin}},
  \bibinfo {author} {\bibfnamefont {V.}~\bibnamefont {Oganesyan}}, \bibinfo
  {author} {\bibfnamefont {J.~M.}\ \bibnamefont {Tranquada}}, \bibinfo {author}
  {\bibfnamefont {A.}~\bibnamefont {Kapitulnik}},\ and\ \bibinfo {author}
  {\bibfnamefont {C.}~\bibnamefont {Howald}},\ }\bibfield  {title} {\bibinfo
  {title} {How to detect fluctuating stripes in the high-temperature
  superconductors},\ }\href {https://doi.org/10.1103/RevModPhys.75.1201}
  {\bibfield  {journal} {\bibinfo  {journal} {Reviews of Modern Physics}\
  }\textbf {\bibinfo {volume} {75}},\ \bibinfo {pages} {1201} (\bibinfo {year}
  {2003})}\BibitemShut {NoStop}%
\bibitem [{\citenamefont {Tranquada}(2015)}]{tranquada_exploring_2015}%
  \BibitemOpen
  \bibfield  {author} {\bibinfo {author} {\bibfnamefont {J.~M.}\ \bibnamefont
  {Tranquada}},\ }\bibfield  {title} {\bibinfo {title} {Exploring intertwined
  orders in cuprate superconductors},\ }\href
  {https://doi.org/10.1016/j.physb.2014.11.056} {\bibfield  {journal} {\bibinfo
   {journal} {Physica B: Condensed Matter}\ }\bibinfo {series} {Special {Issue}
  on {Electronic} {Crystals} ({ECRYS}-2014)},\ \textbf {\bibinfo {volume}
  {460}},\ \bibinfo {pages} {136} (\bibinfo {year} {2015})}\BibitemShut
  {NoStop}%
\bibitem [{\citenamefont {Fradkin}\ \emph {et~al.}(2015)\citenamefont
  {Fradkin}, \citenamefont {Kivelson},\ and\ \citenamefont
  {Tranquada}}]{fradkin_colloquium_2015}%
  \BibitemOpen
  \bibfield  {author} {\bibinfo {author} {\bibfnamefont {E.}~\bibnamefont
  {Fradkin}}, \bibinfo {author} {\bibfnamefont {S.~A.}\ \bibnamefont
  {Kivelson}},\ and\ \bibinfo {author} {\bibfnamefont {J.~M.}\ \bibnamefont
  {Tranquada}},\ }\bibfield  {title} {\bibinfo {title} {\textit{{Colloquium}} :
  {Theory} of intertwined orders in high temperature superconductors},\
  }\href@noop {} {\bibfield  {journal} {\bibinfo  {journal} {Rev. Mod. Phys.}\
  }\textbf {\bibinfo {volume} {87}},\ \bibinfo {pages} {457} (\bibinfo {year}
  {2015})}\BibitemShut {NoStop}%
\bibitem [{\citenamefont {Kivelson}\ \emph
  {et~al.}(1998{\natexlab{a}})\citenamefont {Kivelson}, \citenamefont
  {Fradkin},\ and\ \citenamefont {Emery}}]{Kivelson1998}%
  \BibitemOpen
  \bibfield  {author} {\bibinfo {author} {\bibfnamefont {S.}~\bibnamefont
  {Kivelson}}, \bibinfo {author} {\bibfnamefont {E.}~\bibnamefont {Fradkin}},\
  and\ \bibinfo {author} {\bibfnamefont {V.}~\bibnamefont {Emery}},\ }\bibfield
   {title} {\bibinfo {title} {{Electronic liquid-crystal phases of a doped Mott
  insulator}},\ }\href {https://doi.org/10.1038/31177} {\bibfield  {journal}
  {\bibinfo  {journal} {Nature}\ }\textbf {\bibinfo {volume} {393}},\ \bibinfo
  {pages} {550} (\bibinfo {year} {1998}{\natexlab{a}})},\ \Eprint
  {https://arxiv.org/abs/9707327v1} {arXiv:9707327v1 [arXiv:cond-mat]}
  \BibitemShut {NoStop}%
\bibitem [{\citenamefont {Vojta}\ and\ \citenamefont
  {Sachdev}(1999)}]{Vojta1999}%
  \BibitemOpen
  \bibfield  {author} {\bibinfo {author} {\bibfnamefont {M.}~\bibnamefont
  {Vojta}}\ and\ \bibinfo {author} {\bibfnamefont {S.}~\bibnamefont
  {Sachdev}},\ }\bibfield  {title} {\bibinfo {title} {{Charge Order,
  Superconductivity, and a Global Phase Diagram of Doped Antiferromagnets}},\
  }\href {https://doi.org/10.1103/PhysRevLett.83.3916} {\bibfield  {journal}
  {\bibinfo  {journal} {Phys. Rev. Lett.}\ }\textbf {\bibinfo {volume} {83}},\
  \bibinfo {pages} {3916} (\bibinfo {year} {1999})}\BibitemShut {NoStop}%
\bibitem [{\citenamefont {Sachdev}\ and\ \citenamefont {{La
  Placa}}(2013)}]{Sachdev2013}%
  \BibitemOpen
  \bibfield  {author} {\bibinfo {author} {\bibfnamefont {S.}~\bibnamefont
  {Sachdev}}\ and\ \bibinfo {author} {\bibfnamefont {R.}~\bibnamefont {{La
  Placa}}},\ }\bibfield  {title} {\bibinfo {title} {{Bond order in
  two-dimensional metals with antiferromagnetic exchange interactions}},\
  }\href {https://doi.org/10.1103/PhysRevLett.111.027202} {\bibfield  {journal}
  {\bibinfo  {journal} {Physical Review Letters}\ }\textbf {\bibinfo {volume}
  {111}},\ \bibinfo {pages} {027202} (\bibinfo {year} {2013})},\ \Eprint
  {https://arxiv.org/abs/1303.2114} {arXiv:1303.2114} \BibitemShut {NoStop}%
\bibitem [{\citenamefont {Davis}\ and\ \citenamefont {Lee}(2013)}]{Davis2013}%
  \BibitemOpen
  \bibfield  {author} {\bibinfo {author} {\bibfnamefont {J.~C.~S.}\
  \bibnamefont {Davis}}\ and\ \bibinfo {author} {\bibfnamefont {D.-H.}\
  \bibnamefont {Lee}},\ }\bibfield  {title} {\bibinfo {title} {{Concepts
  relating magnetic interactions, intertwined electronic orders and strongly
  correlated superconductivity}},\ }\bibfield  {journal} {\bibinfo  {journal}
  {Proceedings of the National Academy of Sciences}\ }\textbf {\bibinfo
  {volume} {110}},\ \href {https://doi.org/10.1073/pnas.1316512110}
  {10.1073/pnas.1316512110} (\bibinfo {year} {2013}),\ \Eprint
  {https://arxiv.org/abs/1309.2719} {arXiv:1309.2719} \BibitemShut {NoStop}%
\bibitem [{\citenamefont {Lee}(2014)}]{Lee2014}%
  \BibitemOpen
  \bibfield  {author} {\bibinfo {author} {\bibfnamefont {P.~A.}\ \bibnamefont
  {Lee}},\ }\bibfield  {title} {\bibinfo {title} {{Amperean pairing and the
  pseudogap phase of cuprate superconductors}},\ }\href
  {https://doi.org/10.1103/PhysRevX.4.031017} {\bibfield  {journal} {\bibinfo
  {journal} {Physical Review X}\ }\textbf {\bibinfo {volume} {4}},\ \bibinfo
  {pages} {031017} (\bibinfo {year} {2014})},\ \Eprint
  {https://arxiv.org/abs/1401.0519} {arXiv:1401.0519} \BibitemShut {NoStop}%
\bibitem [{\citenamefont {Wang}\ and\ \citenamefont
  {Chubukov}(2014)}]{Wang2014}%
  \BibitemOpen
  \bibfield  {author} {\bibinfo {author} {\bibfnamefont {Y.}~\bibnamefont
  {Wang}}\ and\ \bibinfo {author} {\bibfnamefont {A.}~\bibnamefont
  {Chubukov}},\ }\bibfield  {title} {\bibinfo {title} {{Charge-density-wave
  order with momentum (2Q,0) and (0,2Q) within the spin-fermion model:
  Continuous and discrete symmetry breaking, preemptive composite order, and
  relation to pseudogap in hole-doped cuprates}},\ }\href
  {https://doi.org/10.1103/PhysRevB.90.035149} {\bibfield  {journal} {\bibinfo
  {journal} {Phys. Rev. B}\ }\textbf {\bibinfo {volume} {90}},\ \bibinfo
  {pages} {35149} (\bibinfo {year} {2014})}\BibitemShut {NoStop}%
\bibitem [{\citenamefont {{Dalla Torre}}\ \emph {et~al.}(2015)\citenamefont
  {{Dalla Torre}}, \citenamefont {He}, \citenamefont {Benjamin},\ and\
  \citenamefont {Demler}}]{DallaTorre2015}%
  \BibitemOpen
  \bibfield  {author} {\bibinfo {author} {\bibfnamefont {E.~G.}\ \bibnamefont
  {{Dalla Torre}}}, \bibinfo {author} {\bibfnamefont {Y.}~\bibnamefont {He}},
  \bibinfo {author} {\bibfnamefont {D.}~\bibnamefont {Benjamin}},\ and\
  \bibinfo {author} {\bibfnamefont {E.}~\bibnamefont {Demler}},\ }\bibfield
  {title} {\bibinfo {title} {{Exploring quasiparticles in high-Tc cuprates
  through photoemission, tunneling, and x-ray scattering experiments}},\ }\href
  {https://doi.org/10.1088/1367-2630/17/2/022001} {\bibfield  {journal}
  {\bibinfo  {journal} {New Journal of Physics}\ }\textbf {\bibinfo {volume}
  {17}},\ \bibinfo {pages} {22001} (\bibinfo {year} {2015})},\ \Eprint
  {https://arxiv.org/abs/1312.0616} {arXiv:1312.0616} \BibitemShut {NoStop}%
\bibitem [{\citenamefont {Duong}\ and\ \citenamefont {Das}(2017)}]{Duong2017}%
  \BibitemOpen
  \bibfield  {author} {\bibinfo {author} {\bibfnamefont {L.~Q.}\ \bibnamefont
  {Duong}}\ and\ \bibinfo {author} {\bibfnamefont {T.}~\bibnamefont {Das}},\
  }\bibfield  {title} {\bibinfo {title} {{Correlation between Fermi arc and
  charge order resulting from the momentum-dependent self-energy correction in
  cuprates}},\ }\href {https://doi.org/10.1103/PhysRevB.96.125154} {\bibfield
  {journal} {\bibinfo  {journal} {Physical Review B}\ }\textbf {\bibinfo
  {volume} {96}},\ \bibinfo {pages} {125154} (\bibinfo {year}
  {2017})}\BibitemShut {NoStop}%
\bibitem [{\citenamefont {Keimer}\ \emph {et~al.}(2015)\citenamefont {Keimer},
  \citenamefont {Kivelson}, \citenamefont {Norman}, \citenamefont {Uchida},\
  and\ \citenamefont {Zaanen}}]{keimer_quantum_2015}%
  \BibitemOpen
  \bibfield  {author} {\bibinfo {author} {\bibfnamefont {B.}~\bibnamefont
  {Keimer}}, \bibinfo {author} {\bibfnamefont {S.~A.}\ \bibnamefont
  {Kivelson}}, \bibinfo {author} {\bibfnamefont {M.~R.}\ \bibnamefont
  {Norman}}, \bibinfo {author} {\bibfnamefont {S.}~\bibnamefont {Uchida}},\
  and\ \bibinfo {author} {\bibfnamefont {J.}~\bibnamefont {Zaanen}},\
  }\bibfield  {title} {\bibinfo {title} {From quantum matter to
  high-temperature superconductivity in copper oxides},\ }\href
  {https://doi.org/10.1038/nature14165} {\bibfield  {journal} {\bibinfo
  {journal} {Nature}\ }\textbf {\bibinfo {volume} {518}},\ \bibinfo {pages}
  {179} (\bibinfo {year} {2015})}\BibitemShut {NoStop}%
\bibitem [{\citenamefont {Radousky}(1992)}]{Radousky1992}%
  \BibitemOpen
  \bibfield  {author} {\bibinfo {author} {\bibfnamefont {H.~B.}\ \bibnamefont
  {Radousky}},\ }\bibfield  {title} {\bibinfo {title} {{A review of the
  superconducting and normal state properties of Y1-xPrxBa2Cu3O7}},\ }\href
  {https://doi.org/10.1557/JMR.1992.1917} {\bibfield  {journal} {\bibinfo
  {journal} {Journal of Materials Research}\ }\textbf {\bibinfo {volume} {7}},\
  \bibinfo {pages} {1917} (\bibinfo {year} {1992})}\BibitemShut {NoStop}%
\bibitem [{\citenamefont {Liang}\ \emph {et~al.}(2006)\citenamefont {Liang},
  \citenamefont {Bonn},\ and\ \citenamefont {Hardy}}]{Liang2006}%
  \BibitemOpen
  \bibfield  {author} {\bibinfo {author} {\bibfnamefont {R.}~\bibnamefont
  {Liang}}, \bibinfo {author} {\bibfnamefont {D.~A.}\ \bibnamefont {Bonn}},\
  and\ \bibinfo {author} {\bibfnamefont {W.~N.}\ \bibnamefont {Hardy}},\
  }\bibfield  {title} {\bibinfo {title} {{Evaluation of CuO2 plane hole doping
  in YBa2 Cu3 O6+x single crystals}},\ }\href@noop {} {\bibfield  {journal}
  {\bibinfo  {journal} {Physical Review B}\ }\textbf {\bibinfo {volume} {73}},\
  \bibinfo {pages} {180505(r)} (\bibinfo {year} {2006})}\BibitemShut {NoStop}%
\bibitem [{\citenamefont {Kivelson}\ \emph
  {et~al.}(1998{\natexlab{b}})\citenamefont {Kivelson}, \citenamefont
  {Fradkin},\ and\ \citenamefont {Emery}}]{kivelson_electronic_1998}%
  \BibitemOpen
  \bibfield  {author} {\bibinfo {author} {\bibfnamefont {S.~A.}\ \bibnamefont
  {Kivelson}}, \bibinfo {author} {\bibfnamefont {E.}~\bibnamefont {Fradkin}},\
  and\ \bibinfo {author} {\bibfnamefont {V.~J.}\ \bibnamefont {Emery}},\
  }\bibfield  {title} {\bibinfo {title} {Electronic liquid-crystal phases of a
  doped {Mott} insulator},\ }\href@noop {} {\bibfield  {journal} {\bibinfo
  {journal} {Nature}\ }\textbf {\bibinfo {volume} {393}},\ \bibinfo {pages}
  {550} (\bibinfo {year} {1998}{\natexlab{b}})}\BibitemShut {NoStop}%
\bibitem [{\citenamefont {Zhao}\ \emph {et~al.}(2019)\citenamefont {Zhao},
  \citenamefont {Ren}, \citenamefont {Rachmilowitz}, \citenamefont
  {Schneeloch}, \citenamefont {Zhong}, \citenamefont {Gu}, \citenamefont
  {Wang},\ and\ \citenamefont {Zeljkovic}}]{zhao_charge-stripe_2019}%
  \BibitemOpen
  \bibfield  {author} {\bibinfo {author} {\bibfnamefont {H.}~\bibnamefont
  {Zhao}}, \bibinfo {author} {\bibfnamefont {Z.}~\bibnamefont {Ren}}, \bibinfo
  {author} {\bibfnamefont {B.}~\bibnamefont {Rachmilowitz}}, \bibinfo {author}
  {\bibfnamefont {J.}~\bibnamefont {Schneeloch}}, \bibinfo {author}
  {\bibfnamefont {R.}~\bibnamefont {Zhong}}, \bibinfo {author} {\bibfnamefont
  {G.}~\bibnamefont {Gu}}, \bibinfo {author} {\bibfnamefont {Z.}~\bibnamefont
  {Wang}},\ and\ \bibinfo {author} {\bibfnamefont {I.}~\bibnamefont
  {Zeljkovic}},\ }\bibfield  {title} {\bibinfo {title} {Charge-stripe crystal
  phase in an insulating cuprate},\ }\href
  {https://doi.org/10.1038/s41563-018-0243-x} {\bibfield  {journal} {\bibinfo
  {journal} {Nature Materials}\ }\textbf {\bibinfo {volume} {18}},\ \bibinfo
  {pages} {103} (\bibinfo {year} {2019})}\BibitemShut {NoStop}%
\bibitem [{\citenamefont {Akhavan}(2002)}]{Akhavan2002}%
  \BibitemOpen
  \bibfield  {author} {\bibinfo {author} {\bibfnamefont {M.}~\bibnamefont
  {Akhavan}},\ }\bibfield  {title} {\bibinfo {title} {{The question of Pr in
  HTSC}},\ }\href {https://doi.org/10.1016/S0921-4526(02)00860-8} {\bibfield
  {journal} {\bibinfo  {journal} {Physica B}\ }\textbf {\bibinfo {volume}
  {321}},\ \bibinfo {pages} {265} (\bibinfo {year} {2002})}\BibitemShut
  {NoStop}%
\bibitem [{\citenamefont {Soderholm}\ \emph {et~al.}(1987)\citenamefont
  {Soderholm}, \citenamefont {Zhang}, \citenamefont {Hinks}, \citenamefont
  {Beno}, \citenamefont {Jorgensen}, \citenamefont {Segre},\ and\ \citenamefont
  {Schuller}}]{Soderholm1987}%
  \BibitemOpen
  \bibfield  {author} {\bibinfo {author} {\bibfnamefont {L.}~\bibnamefont
  {Soderholm}}, \bibinfo {author} {\bibfnamefont {K.}~\bibnamefont {Zhang}},
  \bibinfo {author} {\bibfnamefont {D.~G.}\ \bibnamefont {Hinks}}, \bibinfo
  {author} {\bibfnamefont {M.~A.}\ \bibnamefont {Beno}}, \bibinfo {author}
  {\bibfnamefont {J.~D.}\ \bibnamefont {Jorgensen}}, \bibinfo {author}
  {\bibfnamefont {C.~U.}\ \bibnamefont {Segre}},\ and\ \bibinfo {author}
  {\bibfnamefont {I.~K.}\ \bibnamefont {Schuller}},\ }\bibfield  {title}
  {\bibinfo {title} {{Incorporation of Pr in YBa2Cu3O7-d: electronic effects on
  superconductivity}},\ }\href@noop {} {\bibfield  {journal} {\bibinfo
  {journal} {Nature}\ }\textbf {\bibinfo {volume} {328}},\ \bibinfo {pages}
  {604} (\bibinfo {year} {1987})}\BibitemShut {NoStop}%
\bibitem [{\citenamefont {Neumeier}\ \emph {et~al.}(1989)\citenamefont
  {Neumeier}, \citenamefont {BjÃ¸rnholm}, \citenamefont {Maple},\ and\
  \citenamefont {Schuller}}]{neumeier_hole_1989}%
  \BibitemOpen
  \bibfield  {author} {\bibinfo {author} {\bibfnamefont {J.~J.}\ \bibnamefont
  {Neumeier}}, \bibinfo {author} {\bibfnamefont {T.}~\bibnamefont
  {BjÃ¸rnholm}}, \bibinfo {author} {\bibfnamefont {M.~B.}\ \bibnamefont
  {Maple}},\ and\ \bibinfo {author} {\bibfnamefont {I.~K.}\ \bibnamefont
  {Schuller}},\ }\bibfield  {title} {\bibinfo {title} {Hole filling and pair
  breaking by pr ions in y${}_{1-x}$pr$_{x}$ba$_2$cu$_3$o${}_{6.95 \pm
  0.02}$},\ }\href {https://doi.org/10.1103/PhysRevLett.63.2516} {\bibfield
  {journal} {\bibinfo  {journal} {Physical Review Letters}\ }\textbf {\bibinfo
  {volume} {63}},\ \bibinfo {pages} {2516} (\bibinfo {year} {1989})},\ \bibinfo
  {note} {publisher: American Physical Society}\BibitemShut {NoStop}%
\bibitem [{\citenamefont {Sun}\ \emph {et~al.}(1994)\citenamefont {Sun},
  \citenamefont {Paulius}, \citenamefont {Gajewski}, \citenamefont {Maple},\
  and\ \citenamefont {Dynes}}]{sun_electron_1994}%
  \BibitemOpen
  \bibfield  {author} {\bibinfo {author} {\bibfnamefont {A.~G.}\ \bibnamefont
  {Sun}}, \bibinfo {author} {\bibfnamefont {L.~M.}\ \bibnamefont {Paulius}},
  \bibinfo {author} {\bibfnamefont {D.~A.}\ \bibnamefont {Gajewski}}, \bibinfo
  {author} {\bibfnamefont {M.~B.}\ \bibnamefont {Maple}},\ and\ \bibinfo
  {author} {\bibfnamefont {R.~C.}\ \bibnamefont {Dynes}},\ }\bibfield  {title}
  {\bibinfo {title} {Electron tunneling and transport in the high-$t_c$
  superconductor y${}_{1-x}$pr$_{x}$ba$_2$cu$_3$o${}_{7-\delta}$},\ }\href
  {https://doi.org/10.1103/PhysRevB.50.3266} {\bibfield  {journal} {\bibinfo
  {journal} {Physical Review B}\ }\textbf {\bibinfo {volume} {50}},\ \bibinfo
  {pages} {3266} (\bibinfo {year} {1994})},\ \bibinfo {note} {publisher:
  American Physical Society}\BibitemShut {NoStop}%
\bibitem [{\citenamefont {Frano}\ \emph {et~al.}(2016)\citenamefont {Frano},
  \citenamefont {Blanco-Canosa}, \citenamefont {Schierle}, \citenamefont {Lu},
  \citenamefont {Wu}, \citenamefont {Bluschke}, \citenamefont {Minola},
  \citenamefont {Christiani}, \citenamefont {Habermeier}, \citenamefont
  {Logvenov}, \citenamefont {Wang}, \citenamefont {van Aken}, \citenamefont
  {Benckiser}, \citenamefont {Weschke}, \citenamefont {{Le Tacon}},\ and\
  \citenamefont {Keimer}}]{Frano2016}%
  \BibitemOpen
  \bibfield  {author} {\bibinfo {author} {\bibfnamefont {A.}~\bibnamefont
  {Frano}}, \bibinfo {author} {\bibfnamefont {S.}~\bibnamefont
  {Blanco-Canosa}}, \bibinfo {author} {\bibfnamefont {E.}~\bibnamefont
  {Schierle}}, \bibinfo {author} {\bibfnamefont {Y.}~\bibnamefont {Lu}},
  \bibinfo {author} {\bibfnamefont {M.}~\bibnamefont {Wu}}, \bibinfo {author}
  {\bibfnamefont {M.}~\bibnamefont {Bluschke}}, \bibinfo {author}
  {\bibfnamefont {M.}~\bibnamefont {Minola}}, \bibinfo {author} {\bibfnamefont
  {G.}~\bibnamefont {Christiani}}, \bibinfo {author} {\bibfnamefont {H.~U.}\
  \bibnamefont {Habermeier}}, \bibinfo {author} {\bibfnamefont
  {G.}~\bibnamefont {Logvenov}}, \bibinfo {author} {\bibfnamefont
  {Y.}~\bibnamefont {Wang}}, \bibinfo {author} {\bibfnamefont {P.~A.}\
  \bibnamefont {van Aken}}, \bibinfo {author} {\bibfnamefont {E.}~\bibnamefont
  {Benckiser}}, \bibinfo {author} {\bibfnamefont {E.}~\bibnamefont {Weschke}},
  \bibinfo {author} {\bibfnamefont {M.}~\bibnamefont {{Le Tacon}}},\ and\
  \bibinfo {author} {\bibfnamefont {B.}~\bibnamefont {Keimer}},\ }\bibfield
  {title} {\bibinfo {title} {{Long-range charge-density-wave proximity effect
  at cuprate/manganate interfaces}},\ }\href {https://doi.org/10.1038/nmat4682}
  {\bibfield  {journal} {\bibinfo  {journal} {Nature Materials}\ }\textbf
  {\bibinfo {volume} {15}},\ \bibinfo {pages} {831} (\bibinfo {year}
  {2016})}\BibitemShut {NoStop}%
\bibitem [{\citenamefont {Wojek}\ \emph {et~al.}(2012)\citenamefont {Wojek},
  \citenamefont {Morenzoni}, \citenamefont {Eshchenko}, \citenamefont {Suter},
  \citenamefont {Prokscha}, \citenamefont {Keller}, \citenamefont {Koller},
  \citenamefont {Fischer}, \citenamefont {Malik}, \citenamefont {Bernhard},\
  and\ \citenamefont {D\"obeli}}]{wojek_2012}%
  \BibitemOpen
  \bibfield  {author} {\bibinfo {author} {\bibfnamefont {B.~M.}\ \bibnamefont
  {Wojek}}, \bibinfo {author} {\bibfnamefont {E.}~\bibnamefont {Morenzoni}},
  \bibinfo {author} {\bibfnamefont {D.~G.}\ \bibnamefont {Eshchenko}}, \bibinfo
  {author} {\bibfnamefont {A.}~\bibnamefont {Suter}}, \bibinfo {author}
  {\bibfnamefont {T.}~\bibnamefont {Prokscha}}, \bibinfo {author}
  {\bibfnamefont {H.}~\bibnamefont {Keller}}, \bibinfo {author} {\bibfnamefont
  {E.}~\bibnamefont {Koller}}, \bibinfo {author} {\bibfnamefont
  {O.}~\bibnamefont {Fischer}}, \bibinfo {author} {\bibfnamefont {V.~K.}\
  \bibnamefont {Malik}}, \bibinfo {author} {\bibfnamefont {C.}~\bibnamefont
  {Bernhard}},\ and\ \bibinfo {author} {\bibfnamefont {M.}~\bibnamefont
  {D\"obeli}},\ }\bibfield  {title} {\bibinfo {title} {Magnetism,
  superconductivity, and coupling in cuprate heterostructures probed by
  low-energy muon-spin rotation},\ }\href
  {https://doi.org/10.1103/PhysRevB.85.024505} {\bibfield  {journal} {\bibinfo
  {journal} {Phys. Rev. B}\ }\textbf {\bibinfo {volume} {85}},\ \bibinfo
  {pages} {024505} (\bibinfo {year} {2012})}\BibitemShut {NoStop}%
\bibitem [{SI()}]{SI}%
  \BibitemOpen
  \bibfield  {title} {\bibinfo {title} {{See Supplemental Material at [DOI] for
  Methods and Supplemental Figures}},\ }\href@noop {} {\ }\BibitemShut
  {NoStop}%
\bibitem [{\citenamefont {Fehrenbacher}\ and\ \citenamefont
  {Rice}(1993)}]{Fehrenbacher1993}%
  \BibitemOpen
  \bibfield  {author} {\bibinfo {author} {\bibfnamefont {R.}~\bibnamefont
  {Fehrenbacher}}\ and\ \bibinfo {author} {\bibfnamefont {T.~M.}\ \bibnamefont
  {Rice}},\ }\bibfield  {title} {\bibinfo {title} {{Unusual electronic
  structure of PrBa2Cu3O7}},\ }\href@noop {} {\bibfield  {journal} {\bibinfo
  {journal} {Physical Review Letters}\ }\textbf {\bibinfo {volume} {70}},\
  \bibinfo {pages} {3471} (\bibinfo {year} {1993})}\BibitemShut {NoStop}%
\bibitem [{\citenamefont {Arpaia}\ \emph {et~al.}(2019)\citenamefont {Arpaia},
  \citenamefont {Caprara}, \citenamefont {Fumagalli}, \citenamefont {Vecchi},
  \citenamefont {Peng}, \citenamefont {Andersson}, \citenamefont {Betto},
  \citenamefont {Luca}, \citenamefont {Brookes}, \citenamefont {Lombardi},
  \citenamefont {Salluzzo}, \citenamefont {Braicovich}, \citenamefont {Castro},
  \citenamefont {Grilli},\ and\ \citenamefont
  {Ghiringhelli}}]{arpaia_dynamical_2019}%
  \BibitemOpen
  \bibfield  {author} {\bibinfo {author} {\bibfnamefont {R.}~\bibnamefont
  {Arpaia}}, \bibinfo {author} {\bibfnamefont {S.}~\bibnamefont {Caprara}},
  \bibinfo {author} {\bibfnamefont {R.}~\bibnamefont {Fumagalli}}, \bibinfo
  {author} {\bibfnamefont {G.~D.}\ \bibnamefont {Vecchi}}, \bibinfo {author}
  {\bibfnamefont {Y.~Y.}\ \bibnamefont {Peng}}, \bibinfo {author}
  {\bibfnamefont {E.}~\bibnamefont {Andersson}}, \bibinfo {author}
  {\bibfnamefont {D.}~\bibnamefont {Betto}}, \bibinfo {author} {\bibfnamefont
  {G.~M.~D.}\ \bibnamefont {Luca}}, \bibinfo {author} {\bibfnamefont {N.~B.}\
  \bibnamefont {Brookes}}, \bibinfo {author} {\bibfnamefont {F.}~\bibnamefont
  {Lombardi}}, \bibinfo {author} {\bibfnamefont {M.}~\bibnamefont {Salluzzo}},
  \bibinfo {author} {\bibfnamefont {L.}~\bibnamefont {Braicovich}}, \bibinfo
  {author} {\bibfnamefont {C.~D.}\ \bibnamefont {Castro}}, \bibinfo {author}
  {\bibfnamefont {M.}~\bibnamefont {Grilli}},\ and\ \bibinfo {author}
  {\bibfnamefont {G.}~\bibnamefont {Ghiringhelli}},\ }\bibfield  {title}
  {\bibinfo {title} {Dynamical charge density fluctuations pervading the phase
  diagram of a {Cu}-based high-{Tc} superconductor},\ }\href
  {https://doi.org/10.1126/science.aav1315} {\bibfield  {journal} {\bibinfo
  {journal} {Science}\ }\textbf {\bibinfo {volume} {365}},\ \bibinfo {pages}
  {906} (\bibinfo {year} {2019})}\BibitemShut {NoStop}%
\bibitem [{\citenamefont {Hill}(2000)}]{Hill2000}%
  \BibitemOpen
  \bibfield  {author} {\bibinfo {author} {\bibfnamefont {J.~P.}\ \bibnamefont
  {Hill}},\ }\bibfield  {title} {\bibinfo {title} {{X-ray-scattering study of
  copper magnetism in nonsuperconducting PrBa2Cu3O6.92}},\ }\href
  {https://doi.org/10.7868/s0869565216300150} {\bibfield  {journal} {\bibinfo
  {journal} {Physical Review B}\ }\textbf {\bibinfo {volume} {61}},\ \bibinfo
  {pages} {1251} (\bibinfo {year} {2000})}\BibitemShut {NoStop}%
\bibitem [{\citenamefont {Zhang}\ \emph {et~al.}(2013)\citenamefont {Zhang},
  \citenamefont {Gauquelin}, \citenamefont {Botton},\ and\ \citenamefont
  {Wei}}]{Zhang2013}%
  \BibitemOpen
  \bibfield  {author} {\bibinfo {author} {\bibfnamefont {H.}~\bibnamefont
  {Zhang}}, \bibinfo {author} {\bibfnamefont {N.}~\bibnamefont {Gauquelin}},
  \bibinfo {author} {\bibfnamefont {G.~A.}\ \bibnamefont {Botton}},\ and\
  \bibinfo {author} {\bibfnamefont {J.~Y.}\ \bibnamefont {Wei}},\ }\bibfield
  {title} {\bibinfo {title} {{Attenuation of superconductivity in
  manganite/cuprate heterostructures by epitaxially-induced CuO
  intergrowths}},\ }\href {https://doi.org/10.1063/1.4813840} {\bibfield
  {journal} {\bibinfo  {journal} {Applied Physics Letters}\ }\textbf {\bibinfo
  {volume} {103}},\ \bibinfo {pages} {052606} (\bibinfo {year}
  {2013})}\BibitemShut {NoStop}%
\bibitem [{\citenamefont {Zhang}\ \emph {et~al.}(2018)\citenamefont {Zhang},
  \citenamefont {Gauquelin}, \citenamefont {McMahon}, \citenamefont {Hawthorn},
  \citenamefont {Botton},\ and\ \citenamefont {Wei}}]{Zhang2018}%
  \BibitemOpen
  \bibfield  {author} {\bibinfo {author} {\bibfnamefont {H.}~\bibnamefont
  {Zhang}}, \bibinfo {author} {\bibfnamefont {N.}~\bibnamefont {Gauquelin}},
  \bibinfo {author} {\bibfnamefont {C.}~\bibnamefont {McMahon}}, \bibinfo
  {author} {\bibfnamefont {D.~G.}\ \bibnamefont {Hawthorn}}, \bibinfo {author}
  {\bibfnamefont {G.~A.}\ \bibnamefont {Botton}},\ and\ \bibinfo {author}
  {\bibfnamefont {J.~Y.}\ \bibnamefont {Wei}},\ }\bibfield  {title} {\bibinfo
  {title} {{Synthesis of high-oxidation Y-Ba-Cu-O phases in superoxygenated
  thin films}},\ }\href {https://doi.org/10.1103/PhysRevMaterials.2.033803}
  {\bibfield  {journal} {\bibinfo  {journal} {Physical Review Materials}\
  }\textbf {\bibinfo {volume} {2}},\ \bibinfo {pages} {33803} (\bibinfo {year}
  {2018})}\BibitemShut {NoStop}%
\bibitem [{\citenamefont {Paturi}\ \emph {et~al.}(2004)\citenamefont {Paturi},
  \citenamefont {Peurla}, \citenamefont {Nilsson},\ and\ \citenamefont
  {Raittila}}]{Paturi2004}%
  \BibitemOpen
  \bibfield  {author} {\bibinfo {author} {\bibfnamefont {P.}~\bibnamefont
  {Paturi}}, \bibinfo {author} {\bibfnamefont {M.}~\bibnamefont {Peurla}},
  \bibinfo {author} {\bibfnamefont {K.}~\bibnamefont {Nilsson}},\ and\ \bibinfo
  {author} {\bibfnamefont {J.}~\bibnamefont {Raittila}},\ }\bibfield  {title}
  {\bibinfo {title} {{Crystalline orientation and twin formation in YBCO thin
  films laser ablated from a nanocrystalline target}},\ }\href
  {https://doi.org/10.1088/0953-2048/17/3/043} {\bibfield  {journal} {\bibinfo
  {journal} {Superconductor Science and Technology}\ }\textbf {\bibinfo
  {volume} {17}},\ \bibinfo {pages} {564} (\bibinfo {year} {2004})}\BibitemShut
  {NoStop}%
\bibitem [{\citenamefont {McCoy}\ \emph {et~al.}(2019)\citenamefont {McCoy},
  \citenamefont {Cho}, \citenamefont {Li},\ and\ \citenamefont
  {Cybart}}]{McCoy2019}%
  \BibitemOpen
  \bibfield  {author} {\bibinfo {author} {\bibfnamefont {S.~J.}\ \bibnamefont
  {McCoy}}, \bibinfo {author} {\bibfnamefont {E.~Y.}\ \bibnamefont {Cho}},
  \bibinfo {author} {\bibfnamefont {H.}~\bibnamefont {Li}},\ and\ \bibinfo
  {author} {\bibfnamefont {S.~A.}\ \bibnamefont {Cybart}},\ }\bibfield  {title}
  {\bibinfo {title} {{Ho-Ba-Cu-O Thin Films for Superconductive Electronics}},\
  }\href {https://doi.org/10.1016/S0925-8388(00)00633-2} {\bibfield  {journal}
  {\bibinfo  {journal} {IEEE International Superconductive Electronics
  Conference}\ ,\ \bibinfo {pages} {1}} (\bibinfo {year} {2019})}\BibitemShut
  {NoStop}%
\bibitem [{\citenamefont {Vovk}\ \emph {et~al.}(2013)\citenamefont {Vovk},
  \citenamefont {Nazyrov}, \citenamefont {Goulatis},\ and\ \citenamefont
  {Chroneos}}]{Vovk2013}%
  \BibitemOpen
  \bibfield  {author} {\bibinfo {author} {\bibfnamefont {R.~V.}\ \bibnamefont
  {Vovk}}, \bibinfo {author} {\bibfnamefont {Z.~F.}\ \bibnamefont {Nazyrov}},
  \bibinfo {author} {\bibfnamefont {I.~L.}\ \bibnamefont {Goulatis}},\ and\
  \bibinfo {author} {\bibfnamefont {A.}~\bibnamefont {Chroneos}},\ }\bibfield
  {title} {\bibinfo {title} {{Metal-to-insulator transition in Y1-xPrxBa
  2Cu3O7-$\delta$ single crystals with various praseodymium contents}},\ }\href
  {https://doi.org/10.1016/j.physc.2012.09.017} {\bibfield  {journal} {\bibinfo
   {journal} {Physica C: Superconductivity and its Applications}\ }\textbf
  {\bibinfo {volume} {485}},\ \bibinfo {pages} {89} (\bibinfo {year}
  {2013})}\BibitemShut {NoStop}%
\bibitem [{\citenamefont {Antognazza}\ \emph {et~al.}(1990)\citenamefont
  {Antognazza}, \citenamefont {Triscone}, \citenamefont {Brunner},
  \citenamefont {Mieville}, \citenamefont {Karkut},\ and\ \citenamefont
  {Fischer}}]{Antognazza1990}%
  \BibitemOpen
  \bibfield  {author} {\bibinfo {author} {\bibfnamefont {L.}~\bibnamefont
  {Antognazza}}, \bibinfo {author} {\bibfnamefont {J.~M.}\ \bibnamefont
  {Triscone}}, \bibinfo {author} {\bibfnamefont {O.}~\bibnamefont {Brunner}},
  \bibinfo {author} {\bibfnamefont {L.}~\bibnamefont {Mieville}}, \bibinfo
  {author} {\bibfnamefont {M.~G.}\ \bibnamefont {Karkut}},\ and\ \bibinfo
  {author} {\bibfnamefont {O.}~\bibnamefont {Fischer}},\ }\bibfield  {title}
  {\bibinfo {title} {{Layer by layer preparation of YPBCO alloy thin films}},\
  }\href@noop {} {\bibfield  {journal} {\bibinfo  {journal} {Journal of the
  less common metals}\ }\textbf {\bibinfo {volume} {165}},\ \bibinfo {pages}
  {344} (\bibinfo {year} {1990})}\BibitemShut {NoStop}%
\bibitem [{\citenamefont {Peng}\ \emph {et~al.}(1989)\citenamefont {Peng},
  \citenamefont {Klavins}, \citenamefont {Shelton}, \citenamefont {Radousky},
  \citenamefont {Hahn},\ and\ \citenamefont {Bernardez}}]{Peng1989}%
  \BibitemOpen
  \bibfield  {author} {\bibinfo {author} {\bibfnamefont {J.~L.}\ \bibnamefont
  {Peng}}, \bibinfo {author} {\bibfnamefont {P.}~\bibnamefont {Klavins}},
  \bibinfo {author} {\bibfnamefont {R.~N.}\ \bibnamefont {Shelton}}, \bibinfo
  {author} {\bibfnamefont {H.~B.}\ \bibnamefont {Radousky}}, \bibinfo {author}
  {\bibfnamefont {P.~A.}\ \bibnamefont {Hahn}},\ and\ \bibinfo {author}
  {\bibfnamefont {L.}~\bibnamefont {Bernardez}},\ }\bibfield  {title} {\bibinfo
  {title} {{Upper critical field and normal-state properties of single-phase
  YPBCO compounds}},\ }\href@noop {} {\bibfield  {journal} {\bibinfo  {journal}
  {Physical Review B}\ }\textbf {\bibinfo {volume} {40}},\ \bibinfo {pages}
  {4517} (\bibinfo {year} {1989})}\BibitemShut {NoStop}%
\bibitem [{\citenamefont {Prokscha}\ \emph {et~al.}(2008)\citenamefont
  {Prokscha}, \citenamefont {Morenzoni}, \citenamefont {Deiters}, \citenamefont
  {Foroughi}, \citenamefont {George}, \citenamefont {Kobler}, \citenamefont
  {Suter},\ and\ \citenamefont {Vrankovic}}]{muE4}%
  \BibitemOpen
  \bibfield  {author} {\bibinfo {author} {\bibfnamefont {T.}~\bibnamefont
  {Prokscha}}, \bibinfo {author} {\bibfnamefont {E.}~\bibnamefont {Morenzoni}},
  \bibinfo {author} {\bibfnamefont {K.}~\bibnamefont {Deiters}}, \bibinfo
  {author} {\bibfnamefont {F.}~\bibnamefont {Foroughi}}, \bibinfo {author}
  {\bibfnamefont {D.}~\bibnamefont {George}}, \bibinfo {author} {\bibfnamefont
  {R.}~\bibnamefont {Kobler}}, \bibinfo {author} {\bibfnamefont
  {A.}~\bibnamefont {Suter}},\ and\ \bibinfo {author} {\bibfnamefont
  {V.}~\bibnamefont {Vrankovic}},\ }\bibfield  {title} {\bibinfo {title} {The
  new $\mu$e4 beam at psi: A hybrid-type large acceptance channel for the
  generation of a high intensity surface-muon beam},\ }\href
  {https://doi.org/https://doi.org/10.1016/j.nima.2008.07.081} {\bibfield
  {journal} {\bibinfo  {journal} {Nuclear Instruments and Methods in Physics
  Research Section A: Accelerators, Spectrometers, Detectors and Associated
  Equipment}\ }\textbf {\bibinfo {volume} {595}},\ \bibinfo {pages} {317}
  (\bibinfo {year} {2008})}\BibitemShut {NoStop}%
\bibitem [{\citenamefont {Suter}\ and\ \citenamefont {Wojek}(2012)}]{musrfit}%
  \BibitemOpen
  \bibfield  {author} {\bibinfo {author} {\bibfnamefont {A.}~\bibnamefont
  {Suter}}\ and\ \bibinfo {author} {\bibfnamefont {B.}~\bibnamefont {Wojek}},\
  }\bibfield  {title} {\bibinfo {title} {Musrfit: A free platform-independent
  framework for $\mu$sr data analysis},\ }\href
  {https://doi.org/https://doi.org/10.1016/j.phpro.2012.04.042} {\bibfield
  {journal} {\bibinfo  {journal} {Physics Procedia}\ }\textbf {\bibinfo
  {volume} {30}},\ \bibinfo {pages} {69} (\bibinfo {year} {2012})},\ \bibinfo
  {note} {12th International Conference on Muon Spin Rotation, Relaxation and
  Resonance ($\mu$SR2011)}\BibitemShut {NoStop}%
\bibitem [{\citenamefont {Yaouanc}\ and\ \citenamefont {Dalmas~de
  R\'{e}otier}(2011)}]{yaouanc}%
  \BibitemOpen
  \bibfield  {author} {\bibinfo {author} {\bibfnamefont {A.}~\bibnamefont
  {Yaouanc}}\ and\ \bibinfo {author} {\bibfnamefont {P.}~\bibnamefont
  {Dalmas~de R\'{e}otier}},\ }\href@noop {} {\emph {\bibinfo {title} {Muon Spin
  Rotation, Relaxation, and Resonance}}}\ (\bibinfo  {publisher} {Oxford
  University Press},\ \bibinfo {year} {2011})\BibitemShut {NoStop}%
\end{thebibliography}

%

\end{document}